\documentclass[%
 reprint,
 amsmath,amssymb,
 aps,
]{revtex4-2}

\usepackage{graphicx}
\usepackage{dcolumn}
\usepackage{bm}
\usepackage{dsfont}
\usepackage{subcaption}
\usepackage{caption}
\usepackage{xcolor}

\usepackage{braket}

\begin{document}

\preprint{APS/123-QED}

\title{Topological electronic structures of non-collinear magnetic phases in a multi-orbital Hubbard model with spin-orbit interactions}

\author{Ying-Lin Li$^{1}$}
\author{Po-Hao Chou$^{2,3}$}
\author{Chung-Yu Mou$^{1,2}$}
\affiliation{$^{1}$Center for Quantum Science and Technology and Department of Physics,
National Tsing Hua University, Hsinchu, Taiwan 300, R.O.C.\\ $^{2}$Institute of Physics, Academia Sinica, Nankang, Taiwan, R.O.C. and \\$^{3}$Electrophysics Department, National Yang Ming Chiao Tung University, Hsinchu, Taiwan 300, R.O.C.}%

\begin{abstract}
We explore topological electronic structure of magnetic phases in a multi-orbital Hubbard model with spin-orbit interactions. To account for more general antiferromagnetic orders that go beyond the collinear Néel order, two different spin-orbit interactions, Dresselhaus and Rashba spin-orbit interactions, are considered. By performing the canonical transformation, we derive the corresponding generalized $t$-$J$ model. At half filling, employing self-consistent magnetic order calculations, we find distinctive spin arrangements under Dresselhaus or Rashba spin-orbit interactions. For the Dresselhaus spin-orbit interaction, the spin configuration exhibits collinear antiferromagnetic order. On the other hand, Rashba interaction results in spins antiferromagnetically aligning in  $xy$-plane and  a small interaction controlled by hopping parameter induces spin tilting, causing antiferromagnetic alignment in $xy$-plane but ferromagnetic alignment in $z$-direction. We categorize topological properties of these phases for low doping in the generalized $t$-$J$ model.: for 3D collinear antiferromagnetic order, the system possesses a modified time-reversal symmetry, characterized by the Z$_2$ index. In contrast, for systems with tilted antiferromagnetic orders, it is protected by inversion symmetry and characterized by the Z$_4$ index. We further examine the bulk-edge correspondence for non-collinear magnetic phases, revealing that the surface state becomes gapless when the surface is parallel to the ferromagnetic component of tilted antiferromagnetic order; otherwise, the surface state exhibits a gap. Our findings offer a comprehensive topological characterization for doped and canted antiferromagnetic insulators with spin-orbit interactions, providing valuable insights into the interplay between spin arrangements, symmetries, and topological properties in systems governed by the multi-orbital Hubbard model.
\end{abstract}

\maketitle


\section{\label{sec:introduction}INTRODUCTION}
The search for topological electronic states that are protected by symmetry is one of the foci in recent condensed matter physics. It has led to the discovery of a new class of materials
termed as topological quantum materials, which includes quantum spin Hall insulator\cite{QSHE_PhysRevLett.95.146802, QSHE_PhysRevLett.96.106802, fukanephaffian}, 3D topological insulators\cite{3DTI_PhysRevB.79.195322, 3DTI_PhysRevB.75.121306, 3DTI_PhysRevB.76.045302}, and topological superconductors\cite{TSC_PhysRevB.79.094504, TSC_PhysRevLett.100.096407, TSC_PhysRevLett.102.187001, TSC_PhysRevLett.104.040502, TSC_PhysRevLett.105.097001, TSC_Phys_Rev_B, TSC_communication_physics}.  The original idea for the topological electronic states is that instead of being governed by the conventional order parameter, these electronic structures are governed by non-trivial topology without the presence of conventional orders\cite{QSHE_PhysRevLett.95.146802}.  In particular, for
the generic class - the 3D topological insulator, it is characterized by the macroscopic electrodynamic response with the corresponding action being given by the so-called $\theta$ term
\begin{align}
    S_{\theta} &= \frac{e^{2}}{4\pi hc}\int d^{3}rdt\,  \theta \, \epsilon^{\mu\nu\sigma\tau}\partial_{\mu}A_{\nu}\partial_{\sigma}A_{\tau} \nonumber\\
    &= \frac{e^{2}}{2\pi hc}\int d^{3}rdt \, \theta \, E\cdot B,
\end{align}
where the coefficient $\theta$ is the well-known axion field in the field theory of axion electrodynamics. When the system has time-reversal symmetry, then under the time-reversal transformation $\Theta$, the action should be invariant. Since under the time-reversal transformation, the magnetic field changes sign, $\theta$ is transformed into $-\theta$. Thus $\theta = 0$ or $\pi$, which gives rise to quantized magneto-electric effect.

It was later realized that there are much more topological classes when the conventional order is combined with non-trivial topology. Mong and Moore point it out that in the conventional antiferromagnets, even though the antiferromagnetic order breaks both time-reversal ($\Theta$) and a primitive lattice translational symmetry ($T_{1/2}$) of the crystal,  the combined time-reversal ($\Theta$) and translational ($T_{1/2}$) transformation forms a new symmetry: the $S$-symmetry of the antiferromagnets, and leads to the antiferromagnetic topological insulator (AFTI) phase\cite{Mong_Moore}.  Specifically, because of the $S$-symmetry,  the $\theta$ coefficient in the $\theta$ term is also transformed to $-\theta$\cite{Bernevig} under the $S$ transformation. Thus $\theta$ is also equal to 0 or $\pi$ such that the antiferromagnetic topological insulator still holds quantized magneto-electric effect  in which the non-trivial phase corresponds to the case of $Z_{2} = 1$.


Mong and Moore's work has triggered a number of related  works that study the axion insulator\cite{sekine2021axion_Z4, Turner_Z4, Ono_Z4}. In particular, there has been intensive work trying to realize the AFTI experimentally\cite{Mogi, Zhang_Dongqin, Xu_Yuanfeng, Li_Jiaheng, PhysRevLett.124.066401, Tanaka2020TheoryOI, PhysRevB.106.205107}.  However, to find the AFTI in real materials, one needs to consider realistic magnetic orders by including the spin-orbit interaction and find the corresponding topological characterization. In this case, since the time-reversal symmetry may be broken in the presence of magnetic orders, what is more relevant is the inversion symmetry. As the electric field changes sign under the parity transformation, $\theta$ is transformed into $-\theta$ under the parity transformation. This leads to $\theta = 0$ or $\pi$.  As a result, even in the case when the time-reversal symmetry is broken, the inversion symmetry can still protect the quantized magneto-electric effect\cite{sekine2021axion_Z4}. Thus, in general, an axion insulator refers to a 3D topological insulator, in which the bulk maintains the time-reversal symmetry or inversion symmetry but the surface states may be gapped by surface magnetization\cite{Turner_Z4, Ono_Z4}.  The axion insulator is generally characterized by the $Z_{4}$ index\cite{Ono_Z4}. Specifically,
under the inversion symmetry with translational invariance, in {\it the Brillouin zone defined by antiferromagnetic orders},
the topological index is found to be $(Z_{2})^{3}\times Z_{4}$. Here the $Z_{2}$ index in the $a$ ($a = 1, 2, 3$) direction is given by
 \begin{equation}
        Z_{2} = \prod_{\vec{K} \in \rm{TRIM_a} ; \vec{K} \cdot \hat{a}=0}(-1)^{n^{-}_{\vec{K}}},
 \end{equation}
where TRIM$_a$ represents four time-reversal invariant momenta on $k_a=\pi$ plane and $n_{\vec{K}}^{\pm}$ are number of even-parity and odd-parity eigenvalues of occupied states at 
the TRIM  $\vec{K}$. The product of three $Z_{2}$ indices is a product of $Z_2$ indices for $a=1$, $a=2$, and $a=3$.  The $Z_{4}$ index here is defined as 
    \begin{equation}
    \label{Z4_index}
        Z_{4} = \sum_{\vec{K}\in \rm{TRIMs}} \frac{n_{\vec{K}}^+  - n_{\vec{K}}^-}{2} ({\rm mod} 4),
    \end{equation}
where TRIMs include all time-reversal invariant momenta. Note that  the index for $(Z_2)^3 \times Z_4$ can be expressed
as $(\nu_1, \nu_2, \nu_3; \mu)$ with $\nu_a = 0$ or $1$ and $\mu= 0, 1, 2, 3$.
When $Z_{4}$ is odd (1 or 3),  products of parities for TRIMS on $k_a= 0$ plane and that for TRIMS on  $k_a=\pi$ plane are of opposite sign so that the band gap must vanish somewhere in between\cite{Turner_Z4}. Hence it supports semi-metallic phases such as Weyl semimetallic phase.  As a result, the only topological non-trivial insulating phase is $Z_4 = 2$ phase.  In this case, due to the broken time reversal symmetry, the surface state is gapped. However, the $Z_4=2$ phase is a higher-order topological insulator\cite{Bernevig2017, Xu_Yuanfeng} such that there are protected gapless hinge states at the boundary of surface.

In the previous researches on topological electronic structure in antiferromagnetism\cite{Mogi, Zhang_Dongqin, Xu_Yuanfeng, Li_Jiaheng}, only the spin configurations in which spins are perfectly aligned in the same direction, i.e., the N\'eel orders, are considered. However, in real materials that involve with the spin-orbit interactions, magnetic orders are generally not the N\'eel type orders. It therefore calls for a study of topological electronic states for general antiferromagnetic orders.  In this paper, we investigate real antiferromagnetic spin configurations in a multi-orbital Hubbard model with the Dresselhaus spin-orbit interaction or the Rashba spin-orbit interaction.  We find that for the Dresselhaus spin-orbit interaction, the spin configuration exhibits collinear antiferromagnetic order; while the Rashba interaction results in spins antiferromagnetically aligning in t$xy$-plane but spins are ferromagnetically aligning in $z$-direction.
We categorize the topological properties of these phases. In particular, tilted antiferromagnetic orders are protected by inversion symmetry and described by a Z$_4$ index. Our findings offer a comprehensive topological characterization for canted antiferromagnetic insulators with spin-orbit interactions. This paper is organized as follows: In Sec.\ref{sec:spin arrangement}, we investigate the spin arrangement from a multi-orbital Hubbard mode whose kinetic term is the same as that for the 3D topological insulator. By applying the canonical transformation\cite{tJmoedl_ANDERSON196399, tJmoedl_PhysRevB.14.2989, tJmoedl_PhysRevLett.54.1317, tJmoedl_PhysRevLett.90.207002, tJmoedl_spalek2007tj, tJmoedl_yu2003hubbard, tJmoedl_PhysRevB.94.125135}, we transform the Hubbard model into a multi-orbital spin model.  Based on the derived spin model, we derive the mean field theory and find the self-consistent solution for spin configurations. In Sec.\ref{sec:topological phase},  we first follow Mong and Moore's approach to derive the topological index $Z^M_2$ protected by
the $S$-symmetry on the $k_z=0$ plane at the original Brillouin zone, and then derive the topological index, the $Z_{4}$ index, in the Brillouin zone defined by the antiferromagnetic order. Phase diagrams of the derived multi-orbital spin model are presented. Finally, we conclude and discuss possible extensions of our work in Sec. IV.

\section{\label{sec:spin arrangement}THEORETICAL MODEL AND MEAN-FIELD APPROXIMATION}
We start by considering the Hamiltonian that characterizes a 3D topological insulator in a cubic lattice with either the Dresselhaus interaction $\bar{h}_D$ or the Rashba interaction $\bar{h}_R$\cite{Hosur_Shinsei, Mong_Moore}, which are given by
\begin{eqnarray}
 & &\bar{h}_{D}(k) = {\nu}_{F}\tau^{x}\otimes(\sin{k_{x}}\sigma^{x}+\sin{k_{y}}\sigma^{y}+\sin{k_{z}}\sigma_{z}) \nonumber \\
 & &\hspace{0.5cm}+ [m+t(\cos{k_{x}}+\cos{k_{y}}+\cos{k_{z}})]\tau^{z}, \label{hD}  \\
 & & \bar{h}_{R}(k) = {\nu}_{F}(\tau^{x}\sigma^{y}\sin{k_{x}}-\tau^{x}\sigma^{x}\sin{k_{y}}+\tau^{y}\sin{k_{z}}) \nonumber \\  
 & & \hspace{0.5cm}+ m+t(\cos{k_{x}}+\cos{k_{y}}+\cos{k_{z}})]\tau^{z}. \label{h_R} 
\end{eqnarray}
Here the momentum $k_i$ with $i=x$, $y$, and $z$ are defined by the cubic lattice in the absence of antiferromagnetic order. $\tau_{i}$ and $\sigma_{i}$ are two sets of Pauli matrices that describe the orbit and the spin degrees of freedom respectively, and $v_{F}$ is the Fermi velocity. Note that both $h_R$ and $h_D$ have non-trivial topology in the parameter region $\frac{1}{3}<\frac{t}{m}<1$.  To support magnetic orders,  we further include the multi-orbital Hubbard  interaction $H_U$ and the interactions $H_H$ that realize the Hund's rule\cite{Hund's_1, Hund's_2, Hund's_3}, which are given by
\begin{eqnarray}
& & H_U =  \sum_{i}{U_a n_{i,a\uparrow}n_{i,a\downarrow} + U_{b}n_{i,b\uparrow}n_{i,b\downarrow}}, \\
& & H_H =U'  \sum_{i, \sigma } n_{i a \sigma}n_{i b \Bar{\sigma}} + (U'-J) \sum_{i, \sigma } n_{i a \sigma}n_{i b \sigma} \nonumber \\
& & \hspace{0.8cm}+J \sum_{i} C^{\dagger}_{i a \uparrow}C^{\dagger}_{i b \downarrow}C_{i a \downarrow}C_{i b\uparrow} \nonumber \\
& & \hspace{0.8cm}+J' \sum_{i} C^{\dagger}_{i a \uparrow}C^{\dagger}_{i a \downarrow}C_{i b \downarrow}C_{i b\uparrow}.
\end{eqnarray}
Here the coefficients, $U$ and $U’$ are of similar magnitudes, but $J$ is usually much smaller than $U$. 
In $H_U$, different Hubbard $U$ parameters for two orbits $a$ and $ b$ are introduced and are labeled by $U_a$ and $U_b$. 
In $H_H$,  $\Bar{\sigma} = -\sigma$ and the first and the second terms are the Coulomb interaction between two electrons
in different orbits with opposite and parallel spins. $J$ is the Hund’s exchange coupling. The third term realizes the $x$ and $y$ components of Hund’s exchange that involves spin flipping. The last term is required to insure rotational invariance of interaction so that $U'=U-2J$ and $U>U'>J$.  However, the $J'$-term describes the pair hopping in which two electrons transfer from an orbit to another orbit and is irrelevant to magnetic phases. Hence we shall drop it in the following derivation. Altogether the total Hamiltonian is given by either $H_D = h_D+H_U+H_H$ or $H_R = h_R+H_U+H_H$, where $h_D$ and $h_R$ are the corresponding second quantized Hamiltonians for $\bar{h}_D$ and $\bar{h}_R$ respectively.  As we are interested in magnetic properties, we shall assume that the system is near half filling and in the strong Hubbard $U$ limit. In this region, the spin configuration can be approximated by the spin configuration right at half filling. Since right at half-filling, the ground state is singly occupied at each site for each orbit. By performing the canonical transformation that projects out contribution from doubly occupied states perturbatively in the second order,  the total Hamiltonian will be transformed into a multi-orbital spin model.  In the following, we shall first derive the multi-orbit spin model and the corresponding mean-field equation. 

\subsection{Canonical transformation and the effective spin Hamiltonian}
In large $U_{a,b}$ limit, for each orbit labelled by $\tau$, it is either unoccupied or occupied with spin $\sigma = \uparrow$ or $\sigma = \downarrow$. Hence  the electron operator $C^{\dagger}_{i\tau\sigma}$ can be split as\cite{tJmoedl_PhysRevB.94.125135} 
\begin{equation}
C^{\dagger}_{i\tau\sigma}=C^{\dagger}_{i\tau\sigma}[(1-n_{i\tau\Bar{\sigma}})] + C^{\dagger}_{i\tau\sigma}n_{i\tau\Bar{\sigma}},
\end{equation}
where $\Bar{\sigma} = -\sigma$. Therefore $h_{D,R}$ can be decomposed as 
\begin{equation}
h_{D,R} = \sum_{<i j>, \tau\tau^{\prime}\sigma\sigma^{\prime}}C^{\dagger}_{i\tau\sigma}\hat{T}^{ij}_{\tau\tau^{\prime}\sigma\sigma^{\prime}}C_{j\tau^{\prime}\sigma^{\prime}} + h.c. = \hat{T}_{0} + \hat{T}_{-1} + \hat{T}_{1}, 
\end{equation}
where $\hat{T}^{ij}_{\tau\tau^{\prime}}$ is the corresponding matrix of hopping amplitudes given by $h_R$ or $h_D$ and 
\begin{eqnarray}
&&\hat{T}_{0} = \sum_{<i j>, \tau\tau^{\prime}\sigma\sigma^{\prime}}[(1-n_{i\tau\Bar{\sigma}})C^{\dagger}_{i\tau\sigma}]\hat{T}^{ij}_{\tau\tau^{\prime}\sigma\sigma^{\prime}}[C_{j\tau^{\prime}\sigma^{\prime}}(1-n_{j\tau^{\prime}\Bar{\sigma}})] \nonumber \\
&& +n_{i\tau\Bar{\sigma}}C^{\dagger}_{i\tau\sigma}]\hat{T}^{ij}_{\tau\tau^{\prime}\sigma\sigma^{\prime}}[C_{j\tau\sigma^{\prime}}n_{j\tau\Bar{\sigma}} + h.c. , \\
&&\hat{T}_{-1} = \sum_{<i j>, \tau\tau^{\prime}\sigma\sigma^{\prime}}{(1-n_{i\tau\Bar{\sigma}})C^{\dagger}_{i\tau\sigma} \hat{T}^{ij}_{\tau\tau^{\prime}\sigma\sigma^{\prime}}} C_{j\tau^{\prime}\sigma^{\prime}}n_{j\tau^{\prime}\Bar{\sigma^{\prime}}} \nonumber \\ && + h.c. , \\
 &&\hat{T}_{1} = \sum_{<i j>,\tau\tau^{\prime}\sigma\sigma^{\prime}}{n_{i\tau\Bar{\sigma}}C^{\dagger}_{i\tau\sigma}\hat{T}^{ij}_{\tau\tau^{\prime}\sigma\sigma^{\prime}}C_{j\tau^{\prime}\sigma^{\prime}}(1-n_{j\tau^{\prime}\Bar{\sigma^{\prime}}})} \nonumber \\ &&+ h.c. . 
\end{eqnarray}
Note that $T_{-1}$ and $T_1$ mix doubly occupied space with singly occupied space, while $T_0$ does not.

Let $S$ be the generator of canonical transformation. Then the transformed Hamiltonian $H'$ can be written as
\begin{align}
    \label{eq:H prime}
    H_{D/R}^{\prime}& = e^{iS}H_{D/R} e^{-iS}\nonumber\\
    & = T_{0} + T_{1} + T_{-1} + H_U  + H_H+ [iS, H_{U}]\nonumber\\
    &\hspace{0.5cm}+ [iS,  T_{1} + T_{-1}] + [iS, H_H+T_{0}]+...
\end{align}
To eliminate mixing from doubly occupied space, we require that 
\begin{equation}
[iS, H_{U}] = -(T_{1} + T_{-1}).  \label{condition}
\end{equation}
To find $S$, we consider eigenstates $|m \rangle$ of $H_U$ such that $H_U |m\rangle = E_m |m\rangle$.
Evaluating the matrix element of Eq.(\ref{condition}) in the basis of $|m\rangle$, we have $\langle n | iS, H_{U}] |m \rangle = - \langle n | T_{1} + T_{-1} | m \rangle$. Since $\langle n | iS, H_{U}] |m \rangle = i(E_m-E_n) \langle n | S | m \rangle$, we obtain\cite{tJmoedl_PhysRevB.94.125135}
\begin{equation}
 \langle n | S | m \rangle = i \frac{ \langle n | T_{1} + T_{-1} | m \rangle}{E_m-E_n}.
\end{equation}
Since $T_{-1}$ and $T_1$ connects doubly occupied space with singly occupied space,  for $H_U$ that contains $U$ only for a single orbit, we find that only $E_m-E_n = \pm U$  contributes.
Hence $iS= T_1/U-T_{-1}/U$.  In the case that there are two orbitals in $H_U$, we find that
\begin{equation}
    \label{eq:S operator}
    iS = \sum_{\tau,\tau^{\prime} = a,b;\sigma,\sigma^{\prime}=\uparrow\downarrow}{ \frac{1}{U_{\tau}} (T_1)_{\tau\sigma, \tau^{\prime}\sigma^{\prime}}-\frac{1}{U_{\tau^{\prime}}}(T_{-1})_{\tau\sigma, \tau^{\prime}\sigma^{\prime}}  }
\end{equation}
where $(T_{-1})_{\tau\sigma, \tau^{\prime}\sigma^{\prime}} $ and $(T_1)_{\tau\sigma, \tau^{\prime}\sigma^{\prime}}$ are defined by
\begin{eqnarray}
& & (T_{-1})_{\tau \sigma,\tau^{\prime}\sigma^{\prime}}  = \sum_{i}{(1-n_{i, \tau \bar{\sigma }})C^{\dagger}_{i,\tau \sigma } \hat{T}^{i i+1}_{\tau\tau^{\prime}\sigma\sigma^{\prime}} C_{i+1, \tau^{\prime} \sigma^{\prime}}n_{i+1,\tau^{\prime}\bar{\sigma^{\prime}}}},  \nonumber \\
\label{T_1} \\
& &T_{1,\tau \sigma, \tau^{\prime} \sigma^{\prime}} = \sum_{i}{n_{i, \tau \bar{\sigma }}C^{\dagger}_{i,\tau \sigma }  \hat{T}^{i i+1}_{\tau\tau^{\prime}\sigma\sigma^{\prime}} C_{i+1, \tau^{\prime} \sigma_{\prime}}(1-n_{i+1, \tau^{\prime} \bar{\sigma_{\prime}}})}. \nonumber \\
\label{T_-1}
\end{eqnarray}
Substituting Eq. (\ref{eq:S operator}) into Eq. (\ref{eq:H prime}) and keeping the lowest order term, we find that the effective spin Hamiltonian $H_J$ in low energy sector is given by
\begin{eqnarray}
   && H_J  = H_H+ [iS,  T_{1} + T_{-1}] \nonumber \\
    && = \sum_{\tau,\tau', \sigma, \sigma', \sigma^{\prime\prime}, \sigma^{\prime\prime\prime}}
(\frac{1}{U_{\tau^{\prime}}}+\frac{1}{U_{\tau}})(T_{-1})_{\tau\sigma, \tau^{\prime}\sigma^{\prime}}  (T_1)_{\tau^{\prime}\sigma^{\prime\prime}, \tau\sigma^{\prime\prime\prime}}, \nonumber \\
\end{eqnarray}
where $T_{0}$, $H_{U}$ and $[iS, H_{H}+T_{0}]$ are dropped in low energy sector at half filling without doping so that the resulting Hamiltonian describes interaction between spins. 
To find the effective spin Hamiltonian at half filling, we group the electron $C$ operators into spin operators.
After some algebra, the effective spin Hamiltonian can be transformed to the following forms.
First, when the quadratic term is governed by the Dresselhaus interaction $\bar{h}_D$
\begin{equation*}
\begin{split}
    H^D_J = & \sum_{<ij>} \left\{ v_{F}^{2}(\frac{1}{U_{a}}+\frac{1}{U_{b}})[(S^{x}_{i,a}S^{x}_{j,b}+S^{x}_{i,b}S^{x}_{j,a}) \right. \\
    &\hspace{0.5cm}+(S^{y}_{i,a}S^{y}_{j,b}+S^{y}_{i,b}S^{y}_{j,a})+(S^{z}_{i,a}S^{z}_{j,b}+S^{z}_{i,b}S^{z}_{j,a})\\
        &\hspace{0.5cm}+\frac{3}{4}(n_{i,a}n_{j,b}+n_{i,b}n_{j,a})]\\
        &+\frac{24t^{2}}{U_{a}}(S^{x}_{i,a}S^{x}_{j,a}+S^{y}_{i,a}S^{y}_{j,a}+S^{z}_{i,a}S^{z}_{j,a}-\frac{1}{4}n_{i,a}n_{j,a})\\
        &+\frac{24t^{2}}{U_{b}}(S^{x}_{i,b}S^{x}_{j,b}+S^{y}_{i,b}S^{y}_{j,b}+S^{z}_{i,b}S^{z}_{j,b}-\frac{1}{4}n_{i,b}n_{j,b})\\
        &-2J (S_{i,a}\cdot S_{i,b}+S_{j,a}\cdot S_{j,b})\\
        & \left. +(2U'-J)(\frac{n_{i,a}n_{i,b}}{2}+\frac{n_{j,a}n_{j,b}}{2}) \right \} , 
\end{split}
\end{equation*}
where due to the expectation of the emerging antiferromagnetic interaction,  $\{i,j\}$ labels the sub-lattice in the unit cell of antiferromagnetism and $\{a,b\}$ labels two orbits on each site respectively. Note that we have made use of the following identities: $n_{i, \tau\sigma} + n_{i, \tau\bar{\sigma}} = 1$; $S^{z}_{i\tau} = C^{\dagger}_{i\tau\uparrow}C_{i\tau\uparrow}-C^{\dagger}_{i\tau\downarrow}C_{i\tau\downarrow}$, $S^{+}_{i,\tau} = C^{\dagger}_{i,\tau\uparrow}C_{i, \tau\downarrow}$; $S^{-}_{i,\tau} = C^{\dagger}_{i,\tau\downarrow}C_{i, \tau\uparrow}$; $S^{+}_{i,\tau} = S^{x}_{i,\tau}+iS^{y}_{i,\tau}$ and $S^{-}_{i,\tau} = S^{x}_{i,\tau}-iS^{y}_{i,\tau}$.
On the other hand, when the quadratic term is governed by the Rashba interaction $\bar{h}_R$,
the effective spin Hamiltonian can be transformed to the following form
\begin{equation*}
\begin{split}
    H^R_J = &  \sum_{<ij>} \left\{ v_{F}^{2}(\frac{1}{U_{a}}+\frac{1}{U_{b}})[(S^{x}_{i,a}S^{x}_{j,b}+S^{x}_{i,b}S^{x}_{j,a}) \right. \\
    &\hspace{0.5cm}+(S^{y}_{i,a}S^{y}_{j,b}+S^{y}_{i,b}S^{y}_{j,a})-(S^{z}_{i,a}S^{z}_{j,b}+S^{z}_{i,b}S^{z}_{j,a})\\
        &\hspace{0.5cm}-\frac{3}{4}(n_{i,a}n_{j,b}+n_{i,b}n_{j,a})]\\
        &+\frac{24t^{2}}{U_{a}}(S^{x}_{i,a}S^{x}_{j,a}+S^{y}_{i,a}S^{y}_{j,a}+S^{z}_{i,a}S^{z}_{j,a}-\frac{1}{4}n_{i,a}n_{j,a})\\
        &+\frac{24t^{2}}{U_{b}}(S^{x}_{i,b}S^{x}_{j,b}+S^{y}_{i,b}S^{y}_{j,b}+S^{z}_{i,b}S^{z}_{j,b}-\frac{1}{4}n_{i,b}n_{j,b})\\
        &-2J (S_{i,a}\cdot S_{i,b}+S_{j,a}\cdot S_{j,b})\\
        & \left. +(2U'-J)(\frac{n_{i,a}n_{i,b}}{2}+\frac{n_{j,a}n_{j,b}}{2}) \right\} .
\end{split}
\end{equation*}
Note that the main difference between $H^D_J$ and $H^R_J$ lies in the first term in which the Heisenberg type interaction driven by $v_F$ is no longer antiferromagnetic along $z$-axis in $H^R_J$.

\subsection{\label{sec:mean field result}Magnetic Mean Field Theory}
\begin{figure}[h]
    \begin{subfigure}[b]{0.24\textwidth}
        \includegraphics[width=\textwidth]{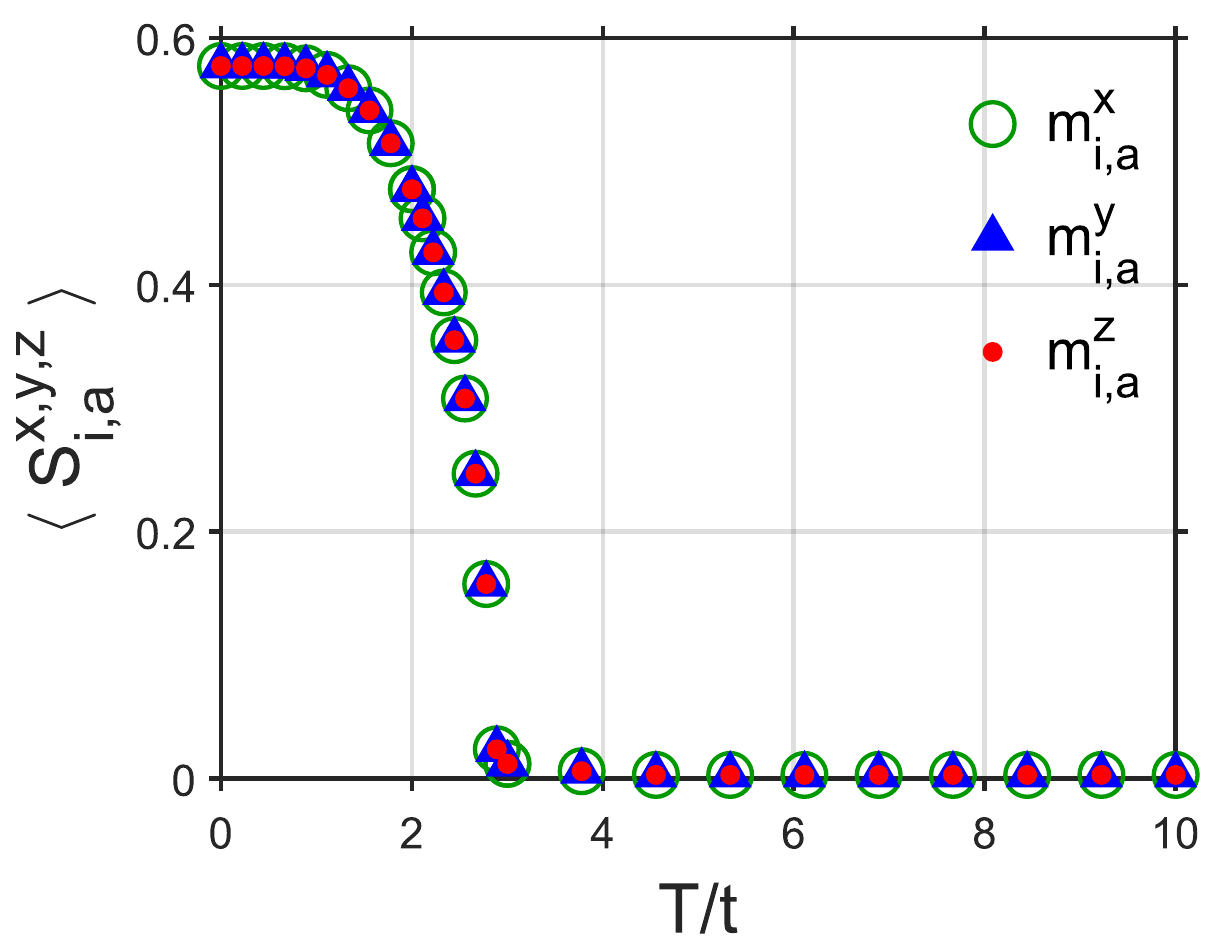}
        \caption{}
    \end{subfigure}
    \begin{subfigure}[b]{0.24\textwidth}
        \includegraphics[width=\textwidth]{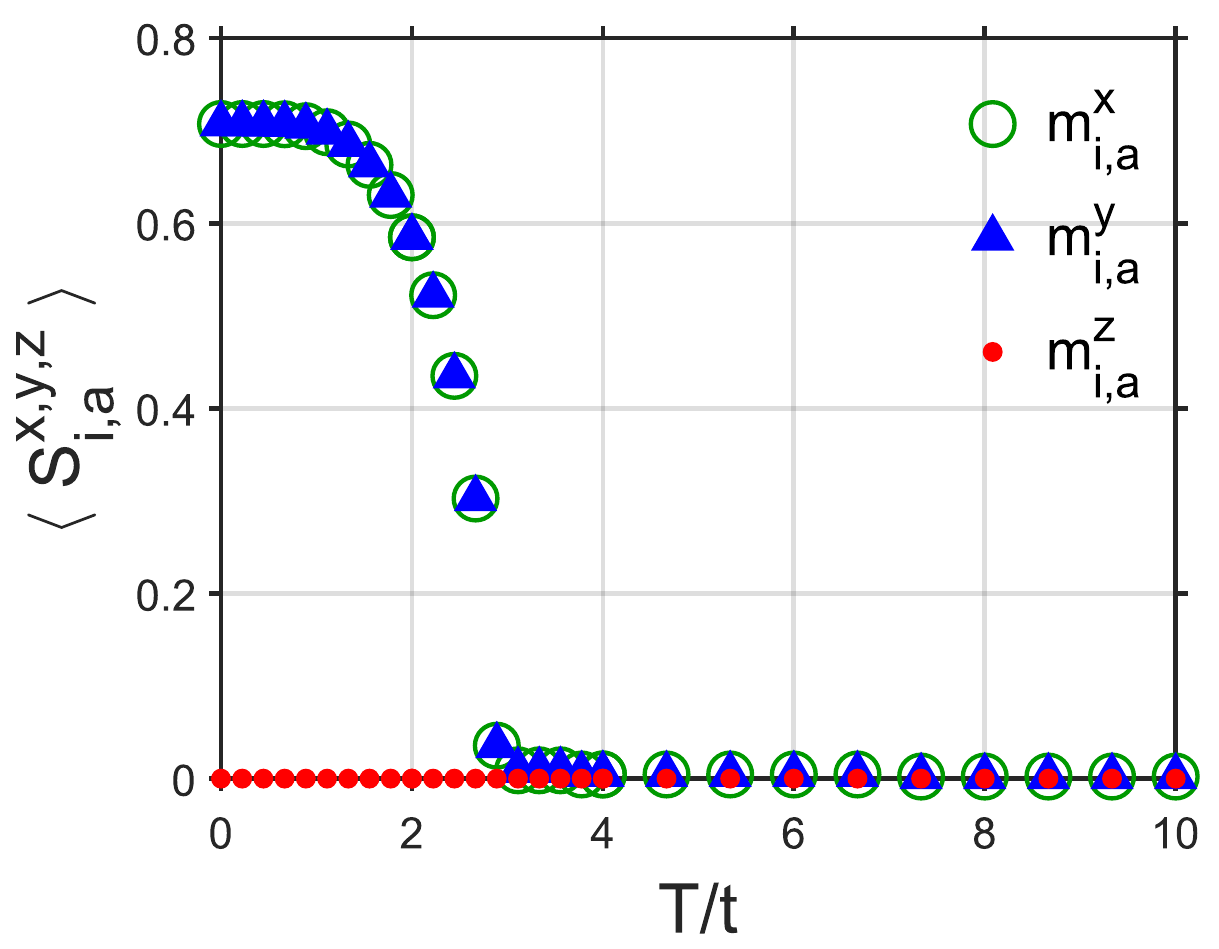}
        \caption{}
    \end{subfigure}
    \begin{subfigure}[b]{0.24\textwidth}
        \includegraphics[width=\textwidth]{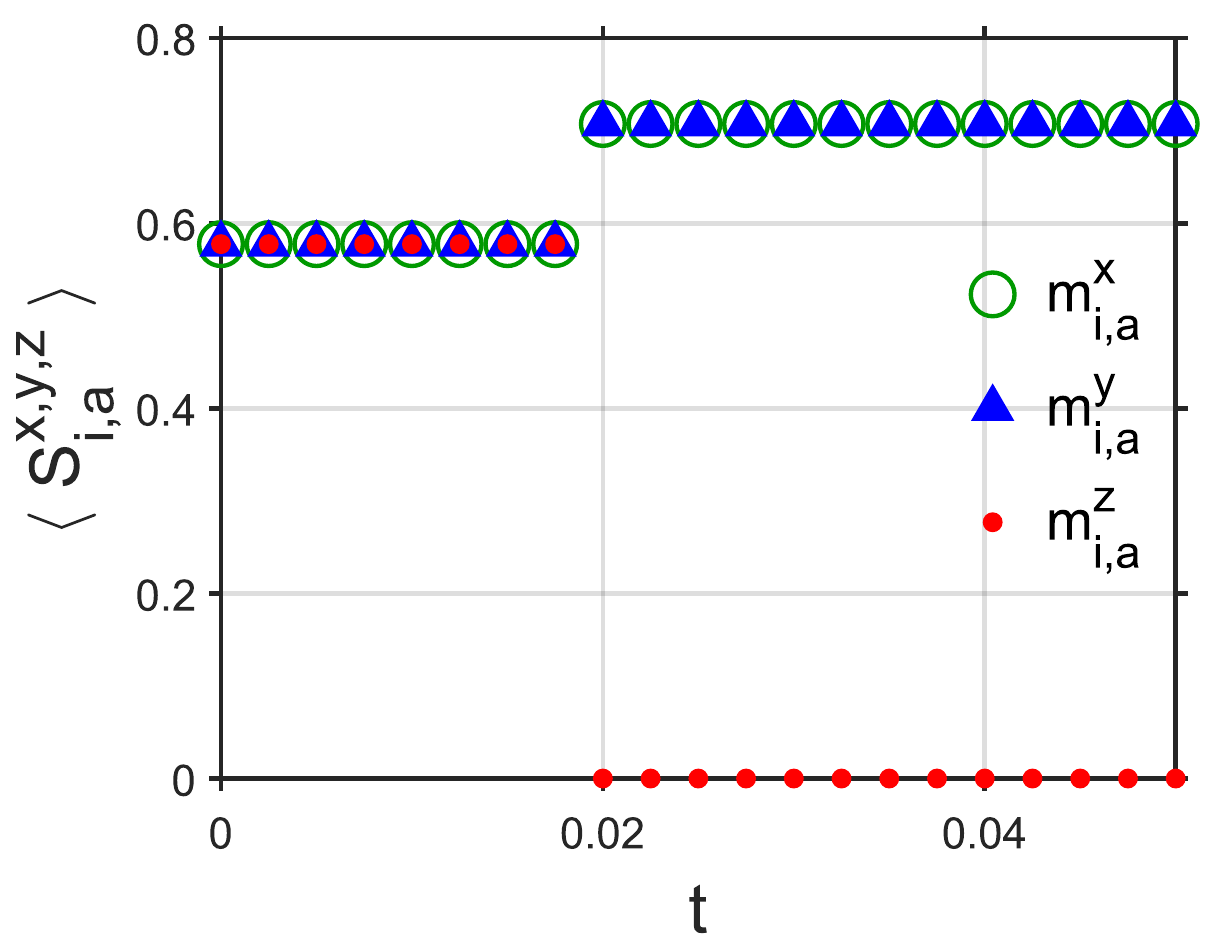}
        \caption{}
    \end{subfigure} 
    \begin{subfigure}[b]{0.24\textwidth}
        \includegraphics[width=\textwidth]{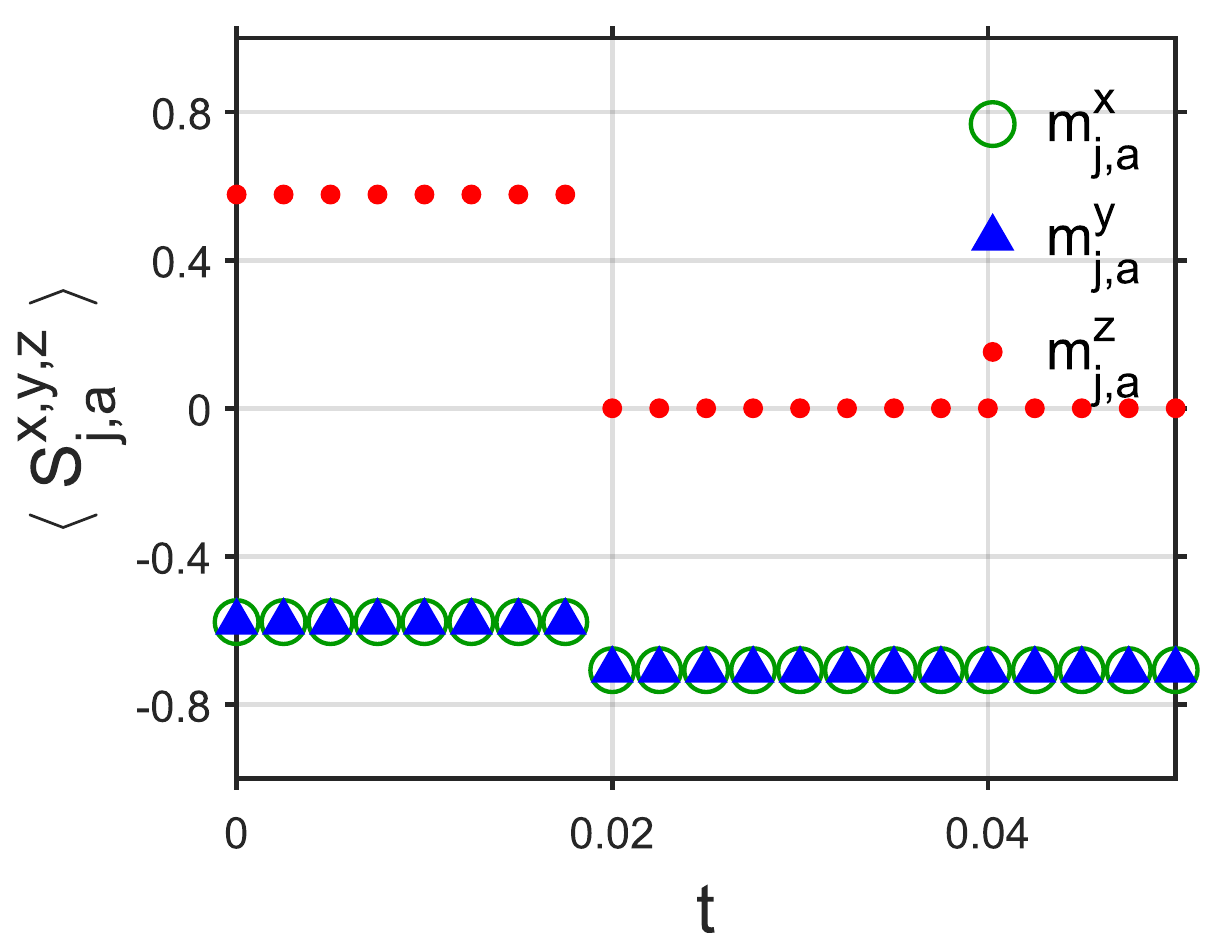}
        \caption{}
    \end{subfigure} 
    
    \begin{subfigure}[b]{0.32\textwidth}
        \includegraphics[width=\textwidth]{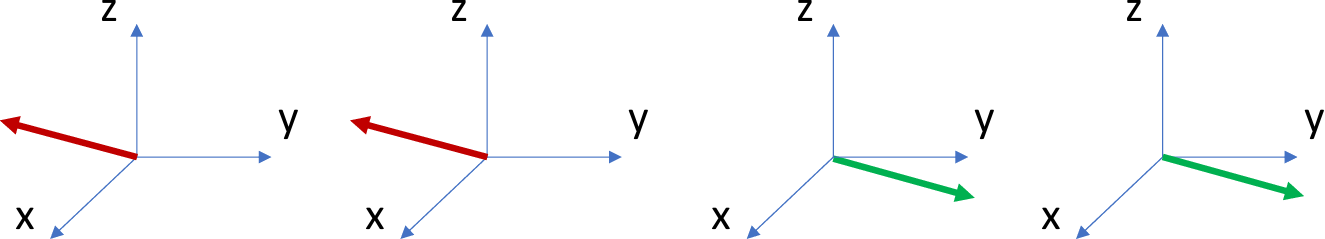}
        \caption{}
    \end{subfigure}
    \begin{subfigure}[b]{0.32\textwidth}
        \includegraphics[width=\textwidth]{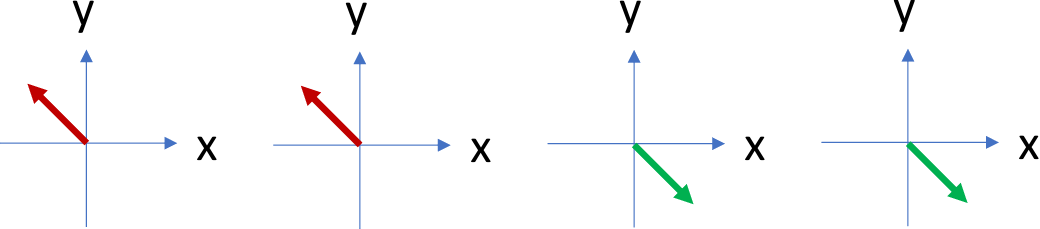}
        \caption{}
    \end{subfigure}
    \begin{subfigure}[b]{0.32\textwidth}
        \includegraphics[width=\textwidth]{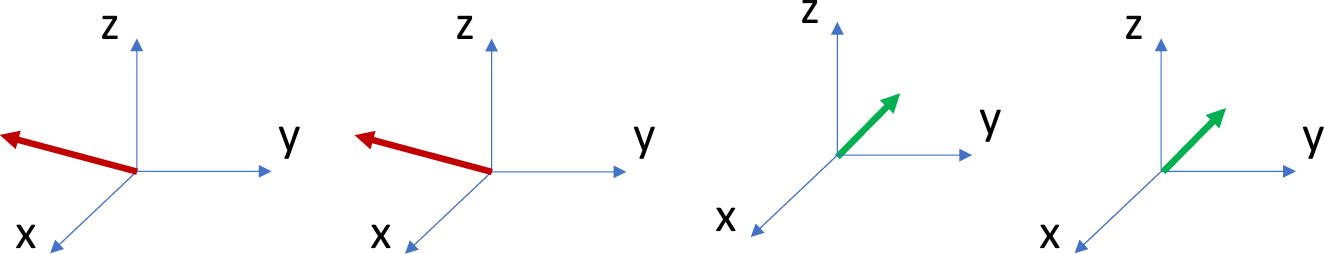}
        \caption{}
    \end{subfigure}
\caption{Mean-field solution for magnetic order parameter  $\langle S^{x,y,z}_{i,\tau}\rangle$:  
Temperature dependence of $\langle S^{x,y,z}_{i,a}\rangle$ for the Hamiltonian with (a) Dresselhaus interaction (b) Rashba interaction.  Here $ m^{x,y,z}_{i,a} = \langle S^{x,y,z}_{i,a}\rangle$, $U_{a}$=$U_{b}$=10, $v_{F}$=0.5, and $J=0.5$. For both cases, $T_{c}/t\approx3$. Magnetic order parameter versus $t$ with $T=0.05$ for the Hamiltonian with the Rashba interaction on  (c) $i$ site  (d)  $j$ site (nearest neighbours to $i$). Here for $t<0.019$, $ m^{x}_{i,a}= m^{y}_{i,a}= m^{z}_{i,a}=0.577$; while $m^{x}_{j,a} = m^{y}_{j,a}= -m^{z}_{j,a}=-0.577$.  For $t \geq 0.019$, the z-components of magnetic orders vanish $m^{z}_{i,a} = m^{z}_{j,a} =0$; while the magnetic orders are antiferromagnetic in $x$-$y$ plane with $m^{x}_{i,a} =m^{y}_{j,a} =- m^{x}_{i,a} = - m^{y}_{j,a} $. Detailed configurations for magnetic orders in orbits $a$ and $b$ for (e) Hamiltonian with the Dresselhaus interaction. Here magnetic orders on the same site point to the same direction for both orbit $a$ and $b$ (red arrows) at $T=0.05$; while magnetic orders on nearest neighbor sites are anti-parallel (green arrows). On the other hand, for Hamiltonian with the Rashba interaction at $T=0.05$,  (f) $z$-component of magnetic orders for $t \geq 0.019$ vanish for both $a$ and $b$ orbits at all sites, while magnetic components on nearest neighbor sites are anti-parallel in the $x$-$y$ plane (red arrows for $i$ site, green arrows for the $j$ site that is a nearest neighbour to $i$.  (g) For $t <0.019$, $z$-component of magnetic orders do not vanish, and  magnetic orders on nearest neighbor sites are antiferromagnetic in the $x$-$y$ plane but are ferromagnetic in the $z$ direction.}
\label{fig:order diagram}
\end{figure}

Once the effective spin Hamiltonian $H_J$ is derived, we can perform the mean-field calculation at half filling.  In the mean approximation, we have
\begin{equation}
 S_{1}\cdot S_{2}\approx\langle S_{2}\rangle\cdot S_{1} + \langle S_{1}\rangle\cdot S_{2} - \langle S_{1}\rangle\cdot\langle S_{2}\rangle,
\end{equation}
the effective Hamiltonian is then converted into $H_{MJ}$.  In this approximation, the mean field partition function is given by 
\begin{equation}
Z= Tr(e^{-\beta H_{MJ}}) =\sum_{E}e^{-\beta E}= e^{-\beta F_{MJ}},
\end{equation} 
where $E$ is an eigenvalue to $H_{MJ}$ and $F_{MJ}$ is the mean-field free energy. The magnetic order parameter is defined by 
\begin{eqnarray}
\langle S_{i\tau}^{x,y,z}\rangle &=& \frac{Tr(S_{i\tau}^{x,y,z}e^{-\beta H_{MJ}})}{Tr(e^{-\beta H_{MJ}})} = \frac{\frac{-2A_{i\tau}^{x,y,z}\sinh(\beta R_{i\tau})}{R_{i\tau}}}{2\cosh(\beta R_{i\tau})} \nonumber \\
       &=& -\frac{A_{i\tau}^{x,y,z}}{R_{i\tau}}\tanh(\beta R_{i\tau}). \label{selfconsistent}
\end{eqnarray}  
Here $\tau =a$ or $b$, $R_{i\tau}= \sqrt{{A_{i\tau}^{x}}^{2}+{A_{i\tau}^{y}}^{2}+{A_{i\tau}^{z}}^{2}}$, and $A_{i \tau}^{x,y,z}$ are given by : \\
For $\tau=a$,
\begin{eqnarray}
&&A_{ia}^{x} =\sum_{j \in <ij>} \nu_{F}^{2}(\frac{1}{U_{a}}+\frac{1}{U_{b}}) \langle S^{x}_{jb}\rangle+ \frac{24t^{2}}{U_{a}} \langle S^{x}_{ja}\rangle -2J\langle S^{x}_{ib}\rangle, \nonumber \\
&&A_{ia}^{y} =\sum_{j \in <ij>} \nu_{F}^{2}(\frac{1}{U_{a}}+\frac{1}{U_{b}}) \langle S^{y}_{jb}\rangle+ \frac{24t^{2}}{U_{a}} \langle S^{y}_{ja}\rangle -2J\langle S^{y}_{ib}\rangle, \nonumber \\
&&A_{ia}^{z} =\sum_{j \in <ij>} \nu_{F}^{2}(\frac{1}{U_{a}}+\frac{1}{U_{b}})\langle S^{z}_{jb}\rangle + \frac{24t^{2}}{U_{a}}\langle S^{z}_{ja}\rangle -2J\langle S^{z}_{ib}\rangle . \nonumber \\
&&
\end{eqnarray}
For $\tau=b$,
\begin{eqnarray}
&&A_{ib}^{x} = \sum_{j \in <ij>} \nu_{F}^{2}(\frac{1}{U_{a}}+\frac{1}{U_{b}}) \langle S^{x}_{ja}\rangle+ \frac{24t^{2}}{U_{a}} \langle S^{x}_{jb}\rangle -2J\langle S^{x}_{ia} \rangle, \nonumber \\
&&A_{ib}^{y} = \sum_{j \in <ij>} \nu_{F}^{2}(\frac{1}{U_{a}}+\frac{1}{U_{b}}) \langle S^{y}_{ja}\rangle+ \frac{24t^{2}}{U_{a}} \langle S^{y}_{jb}\rangle -2J\langle S^{y}_{ia} \rangle, \nonumber \\
&&A_{ib}^{z} = \sum_{j \in <ij>} \nu_{F}^{2}(\frac{1}{U_{a}}+\frac{1}{U_{b}})\langle S^{z}_{ja}\rangle + \frac{24t^{2}}{U_{a}}\langle S^{z}_{jb}\rangle -2J\langle S^{z}_{ia} \rangle . \nonumber \\
&&
\end{eqnarray}
Eq.(\ref{selfconsistent}) form self-consistent equations for magnetic order parameters as the right hand side of the equation depends on the magnetic order parameters through $H_{MF}$.
There are twelve magnetic order parameters which can be chosen to be averaged values for a pair of spins at nearest neighbouring sites $<ij>$: $\langle S^{x,y,z}_{ia}\rangle$,  $\langle S^{x,y,z}_{ib}\rangle$, $\langle S^{x,y,z}_{ja}\rangle$,  $\langle S^{x,y,z}_{jb}\rangle$. We solve these twelve self-consistent equations separately for $H_{MJ}$ with Dresselhaus interaction or Rashba interaction by iteration using 1000 sets of randomly generated initial values of spins. All of the convergent solutions for these 1000 sets initial values show the same pattern of spin configurations as demonstrated in Fig.\ref{fig:order diagram}. Here parameters are taken at typical values: $U_{a}=U_{b}=10$, $v_{F} =0.5$ and $J=0.5$. For all cases, convergent solutions are degenerate in the direction of spin, i.e, if all spins rotated by the same angle,  the configuration is still a solution to Eq.(\ref{selfconsistent}). This property reflects rotational symmetry of $H^{D/R}_J$. In Figs.\ref{fig:order diagram}(a) and (b), we show typical convergent average of spin components versus temperature over $t$ with $t=1$ for the Hamiltonian with the Dresselhaus interaction or the  Rashba interaction. Clearly, it is seen that  for both cases, $T_{c}/t\approx3$.  In particular, for the Hamiltonian with the Dresselhaus interaction, the magnetic order parameter always has non-vanishing 3 components; while for  the Hamiltonian with the Rashba interaction, the $z$ component of the magnetic order vanishes for $t=1$. Indeed, in Figs.\ref{fig:order diagram}(c) and (d), we show convergent magnetic order parameters at two nearest neighbouring sites $i,j$ versus $t$ with $T=0.05$ for the the Hamiltonian with the Rashba interaction.  It is seen that for $t \geq 0.019$, the $z$ component of the magnetic order vanishes; while for $t < 0.019$, $z$ components of the magnetic orders at two nearest neighbouring sites $i$ and $j$ show non-vanishing but different values. The detailed configurations for magnetic orders in orbits $a$ and $b$ are shown in Figs. \ref{fig:order diagram}(e), (f), and (g). Here as shown in Fig. \ref{fig:order diagram}(e) for the Hamiltonian with the Dresselhaus interaction, spins on the same site point to the same direction for both orbit $a$ and $b$; while spins on nearest neighbor sites are anti-parallel.  On the other hand, for the Hamiltonian with the Rashba interaction at $T=0.05$, $z$-component of magnetic orders for $t \geq 0.019$ vanish for both $a$ and $b$ orbits at all sites, while magnetic components on nearest neighbor sites are anti-parallel in the $x$-$y$ plane as shown in 
Fig. \ref{fig:order diagram}(f). However, for $t <0.019$, $z$-component of magnetic orders do not vanish. As shown in Fig. \ref{fig:order diagram}(g), we find that magnetic orders on nearest neighbor sites are antiferromagnetic in the $x$-$y$ plane but are ferromagnetic in the $z$ direction. These results are expected to hold qualitatively correct in configuration of directions even if we go beyond the mean-field theory and include effects of fluctuations.
\onecolumngrid

\twocolumngrid
\section{\label{sec:topological phase}TOPOLOGICAL PHASES UNDER SPIN-ORBIT INTERACTION}
Based on the magnetic mean-field at half filling obtained in sec.\ref{sec:mean field result}, we shall explore the topological electronic structure at low hole-doing region.
This will be done by combining $h_D$ ot $h_R$ with $H^D_J$ or $H^R_J$ appropriately.  In this work, we shall consider hole doping characterized by $\delta$.  In the large $U$ limit, the effect of no double occupancy renormalizes $h_{D/R}$ and $H^{D/R}_J$.  It is shown that the renormalization can be included as renormalized factor $g_t=2\delta/(1+\delta)$ and $g_J=4/(1+\delta)^2$ such that the total renormalized Hamiltonians are given by\cite{Zhang}
\begin{equation}
    H^{D/R} = g_t h_{D/R} + g_J H^{D/R}_J.
\end{equation}
The corresponding renormalized mean-field Hamiltonians are then given by
\begin{equation}
    H^{D/R}_M = g_t h_{D/R} + g_J H^{D/R}_{MJ}.
\end{equation}
In the following, we shall find $H^{D/R}_M$ explicitly and find the corresponding electronic topological structures.

\subsection{\label{sec:RMFT}Renormalized Mean-field Theory in Low Doping Region}
The renormalized mean-field Hamiltonian $H^{D/R}_M$ is quadratic and can be generally written as $H^{D/R}_M = \sum_k \psi^{\dagger} (k) h_{D/R} (k) \psi(k)$ with $\psi(k)$ being the corresponding electron operator in $k$ space. We shall first find $ h_{D/R} (k)$ for a given doping $\delta$.
For this purpose, we first note that the cubic lattice is bipartite with $A$ and $B$ sub-lattices, hence in the presence of antiferromagnetic order, the unit cell is composed by a pair of nearest $A$ and $B$ lattice points and the lattice vectors that connect unit cells are $a_1=(1,0,1)$, $a_2=(0,1,1)$, and $a_3=(0,0,2)$.  Therefore, to accommodate the antiferromagnetic order, we need to double the matrix size of the Hamiltonian. Following the same procedure as Mong et al.\cite{Mong_Moore} did, we first decompose any Hamiltonian $h(k)$ in the original $k$ space in the following form
\begin{equation}
    h(k_x, k_y, k_z) = T^0 + \sum_{\alpha= x,y,z}{(T_{\alpha}e^{-ik_{\alpha}} + T^{\dagger}_{\alpha}e^{k_{\alpha}})},
\end{equation}
where $T_{\alpha}$ describes hopping from neighbouring sites from the $-\alpha$ direction, $T^{\dagger}_{\alpha}$ are hopping from $+\alpha$ direction, and $T^0$ describes the self-interaction on the same site.  For the Hamiltonian with the Dresselhaus interaction, the antiferromagnetic order has three non-vanishing components. The enlarged eight-band Hamiltonian is given by
\onecolumngrid
\begin{equation*}
    h(k_x, k_y, k_z) = 
\begin{bmatrix}
T^{0}+M & 0\\
0 & T^{0}-M
\end{bmatrix}
    + \sum_{\alpha=x,y,z}
\begin{bmatrix}
0 & (T_{\alpha}e^{-ik_{\alpha}}+T_{\alpha}^{\dagger}e^{ik_{\alpha}})e^{ik_z}\\
(T_{\alpha}e^{-ik_{\alpha}}+T_{\alpha}^{\dagger}e^{ik_{\alpha}})e^{-ik_z} & 0
\end{bmatrix} - \mu I .
\end{equation*}
\twocolumngrid
\noindent Here $\mu$ is the chemical potential determined by doping. However, to explore possible topological electronic structures, it is convenient to set it to zero.
We still describe the Hamiltonian by the original cubic lattice's basis and the matrix$\left( \begin{smallmatrix} M & 0 \\ 0 & -M \end{smallmatrix} \right)$ describes the antiferromagnetic N\'eel order in $A$ and $B$ sub-lattices along $z$ direction with 
$M = \sum_{\alpha=x,y,z} M_{AF,\alpha}1_{\tau}\otimes\sigma^{\alpha}$ being a $4 \times 4$ matrix characterizing magnetic order $(M_{AF,x},M_{AF,y},M_{AF,z})$ in $A$ sub-lattice.
Without loss of generality, we shall normalized the magnitude of ${\vec{M}}_{AF}$ to one, i.e., set $M_{AF,x}^2+M_{AF,y}^2+M_{AF,z}^2=1$, in the following analysis.  The renormalized hopping matrices are given by $T^{0} =  \bar{m} \tau^{z}\otimes1_{\sigma}$ and $T_{\alpha} = \frac{1}{2} \bar{t} \tau^{z}\otimes1_{\sigma} + \frac{1}{2} i \bar{v}_{F}\tau^{x}\otimes\sigma^{\alpha}$ with 
\begin{eqnarray} \label{bar}
&&\bar{t}= g_t t = \frac{2\delta}{1+\delta} \times t, \nonumber \\
&&\bar{m}= g_t m = \frac{2\delta}{1+\delta} \times m , \nonumber \\
&& \bar{v}_F= g_t v_F = \frac{2\delta}{1+\delta} \times v_F.
\end{eqnarray}
After some algebra, the renormalized mean-field Hamiltonian with Dresselhaus interaction and mean-field antiferromagnetic order is given by
\begin{eqnarray} 
    \label{hD_AF}
     && h_D(k) =  \bar{m}  \tau^{z} + \sum_{i=x, y, z} g_J M_{AF,i} \gamma^{z}\sigma^{i} \nonumber\\   
     &&+ \bar{t} \cos{k_{x}}\cos{k_{z}} \gamma^{x}\tau^{z} +  \bar{t} \cos{k_{x}}\sin{k_{z}} \gamma^{y}\tau^{z}\nonumber\\
     &&+ \bar{v}_{F} \sin{k_{x}}\cos{k_{z}} \gamma^{x}\tau^{x}\sigma^{x} + \bar{v}_{F} \sin{k_{x}}\sin{k_{z}} \gamma^{y}\tau^{x}\sigma^{x}\nonumber\\   
     && + \bar{t} \cos{k_{y}}\cos{k_{z}} \gamma^{x}\tau^{z} + \bar{t} \cos{k_{y}}\sin{k_{z}} \gamma^{y}\tau^{z}\nonumber\\    
     && + \bar{v}_{F} \sin{k_{y}}\cos{k_{z}} \gamma^{x}\tau^{x}\sigma^{y} + \bar{v}_{F} \sin{k_{y}}\sin{k_{z}} \gamma^{y}\tau^{x}\sigma^{y}\nonumber\\    
     && + \bar{t} (\cos k_{z})^{2} \gamma^{x}\tau^{z} + \bar{t} \cos{k_{z}}\sin{k_{z}} \gamma^{y}\tau^{z}\nonumber\\    
     && +\bar{v}_{F} \sin{k_{z}}\cos{k_{z}} \gamma^{x}\tau^{x}\sigma^{z} + \bar{v}_{F} (\sin k_{z})^{2} \gamma^{y}\tau^{x}\sigma^{z},
\end{eqnarray}
where the Pauli matrices, $\gamma^{x,y,z}$, are introduced to describe antiferromagnetic bipartite degree of freedom.
By the same procedure applied for the Hamiltonian with the Rashba interaction, the renormalized mean-field Hamiltonian is given by
\begin{align}
    \label{h_R_AFxy_Fz}
    h_R(k) = & \bar{m} \tau^{z} +g_J M_{AF,x} \gamma^{z}\sigma^{x} + g_J M_{AF,y} \gamma^{z}\sigma^{y} + g_J M_{F,z} \sigma^{z}\nonumber\\
    & + \bar{t} \cos{k_{x}}\cos{k_{z}} \gamma^{x}\tau^{z} + \bar{t} \cos{k_{x}}\sin{k_{z}} \gamma^{y}\tau^{z}\nonumber\\
    & + \bar{v}_{F} \sin{k_{x}}\cos{k_{z}} \gamma^{x}\tau^{x}\sigma^{y} + \bar{v}_{F} \sin{k_{x}}\sin{k_{z}} \gamma^{y}\tau^{x}\sigma^{y}\nonumber\\
    & + \bar{t} \cos{k_{y}}\cos{k_{z}} \gamma^{x}\tau^{z} + \bar{t} \cos{k_{y}}\sin{k_{z}} \gamma^{y}\tau^{z}\nonumber\\
    & + \bar{v}_{F} \sin{k_{y}}\cos{k_{z}} \gamma^{x}\tau^{x}\sigma^{x} + \bar{v}_{F} \sin{k_{y}}\sin{k_{z}} \gamma^{y}\tau^{x}\sigma^{x}\nonumber\\
    & + \bar{t} (\cos k_{z})^{2} \gamma^{x}\tau^{z} + \bar{t} \cos{k_{z}}\sin{k_{z}} \gamma^{y}\tau^{z}\nonumber\\
    & + \bar{v}_{F} \sin{k_{z}}\cos{k_{z}} \gamma^{x}\tau^{y} + \bar{v}_{F} (\sin k_{z})^{2} \gamma^{y}\tau^{y},
\end{align}
where $M_{F,z} \geq 0 $ and it is the ferromagnetic term in $z$-direction. This model includes two situations, $M_{F,z} = 0$ and antiferromagnetic ordering in $x$-$y$ plane and antiferromagnetic ordering in $x$-$y$ plane and ferromagnetic ordering in $z$ direction.

\subsection{\label{sec:topology} Topological Phases in Low Doping Region}
In this subsection, we shall first follow Mong and Moore's approach to derive the topological index $Z^M_2$ protected by
the $S$-symmetry on the $k_z=0$ plane in the original Brillouin zone, and then derive the topological index, the $Z_{4}$ index, in the Brillouin zone defined by the antiferromagnetic order.
First, as shown previously\cite{Mong_Moore,sekine2021axion_Z4, Shiozaki2016TNCI}, the two renormalized mean-field  Hamiltonian, Eq.(\ref{hD_AF}) and Eq.(\ref{h_R_AFxy_Fz}) with $M_{F,z}=0$, are protected by the effective time-reversal symmetry $S$ that combines the time-reversal symmetry and a lattice translation. More precisely, both Hamiltonians satisfies $Sh(k)S^{-1}=h(-k)$ with $S(k) = \Theta T_{1/2}(k)$. Here  $\Theta$ is the time-reversal symmetry operator given by 
\begin{equation*}
    \Theta = -i
    \begin{bmatrix}
    1_{\tau}\otimes\sigma^{y} & 0\\
    0 & 1_{\tau}\otimes\sigma^{y}
    \end{bmatrix}
    \kappa
\end{equation*}
with $\kappa$ being the operator that takes complex conjugation; while $T_{1/2}(k)$ is a translation though half of a unit cell, given by
\begin{equation*}
    T_{1/2}(k) = e^{ik_{z}}
    \begin{bmatrix}
    0 & 1_{\tau}\otimes1_{\sigma}\\
    e^{-2ik_{z}}1_{\tau}\otimes1_{\sigma} & 0
    \end{bmatrix}.
\end{equation*}
Clearly, at the plane $k_z=0$, one finds $[S({k_x,k_y, 0})]^2=-1$. Hence by analogy to the $Z_2$ invariance in
the quantum-spin-Hall effect, one can apply the $Z_2$ topological classification that works for 2D systems to the 3D antiferromagnetic system in the Brillouin zone at the plane $k_{z}=0$. Here  the $Z_2$ invariant can be computed at the plane $k_z=0$\cite{Mong_Moore}, to distinguish it from the usual $Z_2$ index, we shall denote it by $Z^M_2$. Note that $[S(k_x,k_y, \pi)]^2=+1$ so that there is no topological invariant associated with the plane $k_z=\pi$. In addition, in presence of inversion symmetry, the $Z_2$ topological invariant can be computed at the four time-reversal invariant momenta with $k_x$ and $k_y$ being either $0$ or $\pi$\cite{3DTI_PhysRevB.76.045302}. For the two renormalized mean-field Hamiltonians, Eq.(\ref{hD_AF}) and Eq.(\ref{h_R_AFxy_Fz}), we find that both $h_D$ and $h_R$ preserve the $k$-dependent inversion symmetry with the transformation being defined by
\begin{equation}
\label{inversion_symmetry}
    I = 
    \begin{bmatrix}
        1 & 0\\
        0 & e^{-2i k_{z}}
    \end{bmatrix}
    \tau^{z}.
\end{equation}

To find the $Z^M_2$ invariant for our 3D antiferromagnetic insulator in the presence of inversion symmetry, we follow Refs.\cite{fukanephaffian, 3DTI_PhysRevB.76.045302} and compute the  $Z^M_{2}$ index $\nu$ as
\begin{equation}
\label{eq:z2_pf}
    (-1)^{\nu} = \prod_{i}^{4}{\frac{\sqrt{\det[w(\Gamma_i)]}}{{\rm pf}[w(\Gamma_i)]}}.
\end{equation}
Here when $\nu = 1$, the system is in non-trivial phase. $\Gamma_i$ ($i=1,2,3,4$) are four time reversal invariant points $(0,0,0), (\pi,0,0), (\pi,\pi,0)$ and $(0,\pi,0)$. $w$ is a $2N \times 2N$ matrix with $2N$ being the number of occupied states. The matrix element of $w$ is defined as $w_{mn} = \bra{u_{m}(-k)}\Theta\ket{u_{n}(k)}$ with $u_{n}$ being the $n_{th}$ occupied state. At  the time reversal invariant momentum $\Gamma_i$,  $w_{mn}$ is a skew-symmetric matrix so the Pfaffian ${\rm pf} [w]$ can be found through the relation ${\rm pf} [w]^2= \det (w)$.  Clearly, $\sqrt{\det w}$ and ${\rm pf} [w]$ differ by sign such that $\sqrt{\det w}/{\rm pf} [w] = \pm 1$. In the presence of inversion symmetry, these signs determine the topological index through the following relation
\begin{equation}
 \frac{\sqrt{\det[w(\Gamma_i)]}}{{\rm pf}[w(\Gamma_i)]} = \prod_{m=1}^{N}\xi_{2m}(\Gamma_i),
\end{equation}
where $\xi_{2m}(\Gamma_i) = +1$ or $-1$ represents the parity eigenvalue of occupied state $ | u_{2m} \rangle$ at $\Gamma_{{i}}$.  Note that due to the effective time-reversal symmetry, the degenerate Kramer's pairs share the same eigenvalue, $\xi_{2m} = \xi_{2m-1}$, hence we only need to consider even or odd occupied states in the parity calculation.
 
\begin{center}
\begin{figure}[h]
    \begin{subfigure}[b]{0.44\textwidth}
        \centering        
        \includegraphics[width=\textwidth]{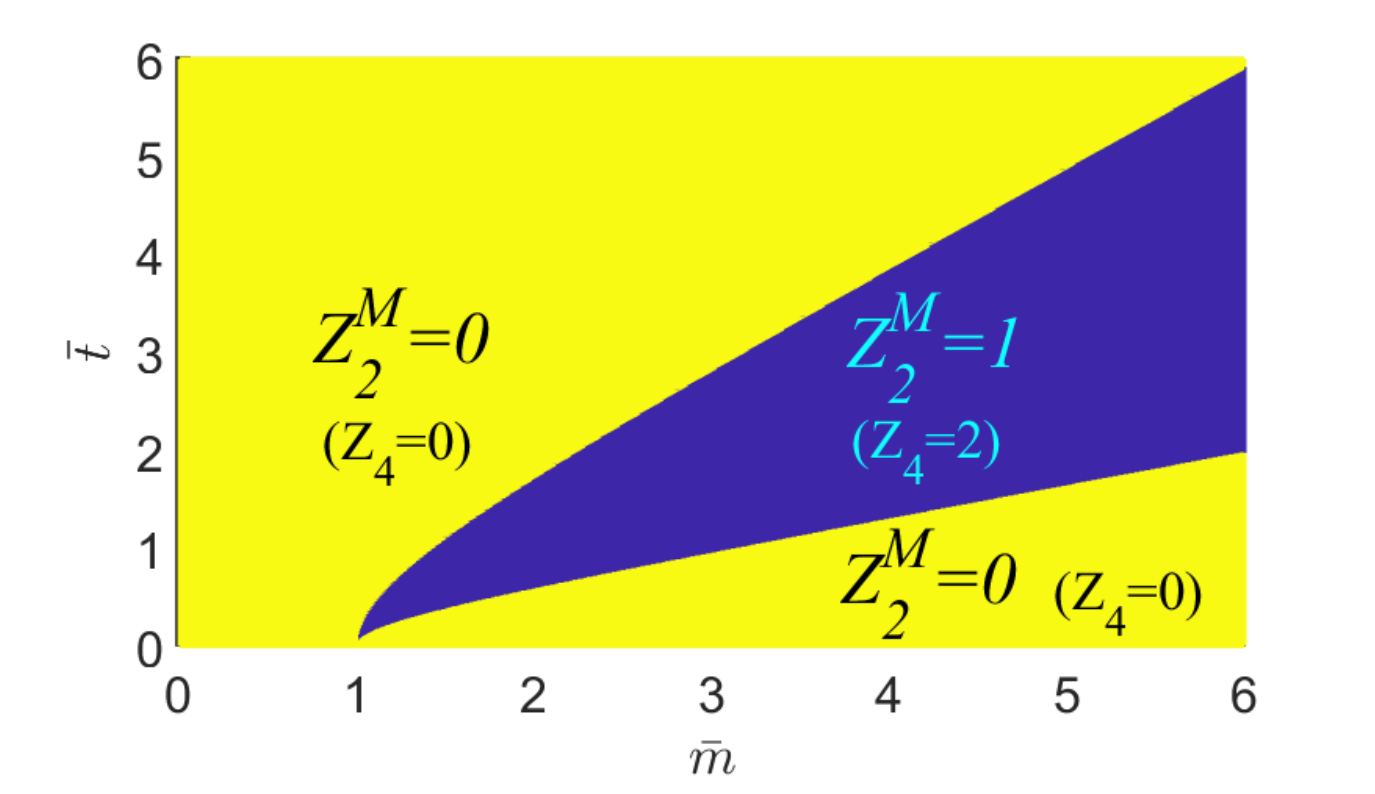}
        \caption{}
    \end{subfigure}
    \begin{subfigure}[b]{0.44\textwidth}
        \centering        
        \includegraphics[width=\textwidth]{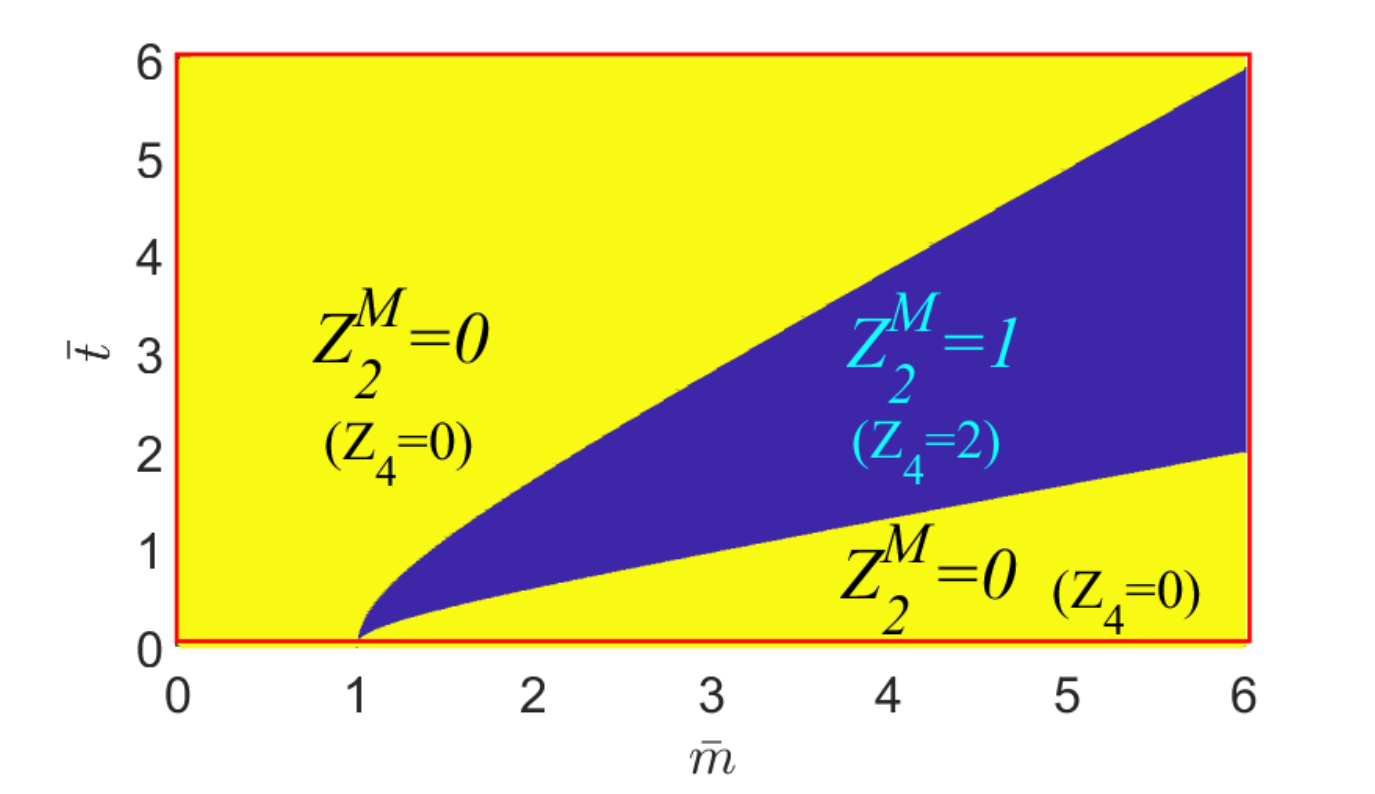}
        \caption{}
    \end{subfigure}

    \begin{subfigure}[b]{0.4\textwidth}
        \centering        
        \includegraphics[width=\textwidth]{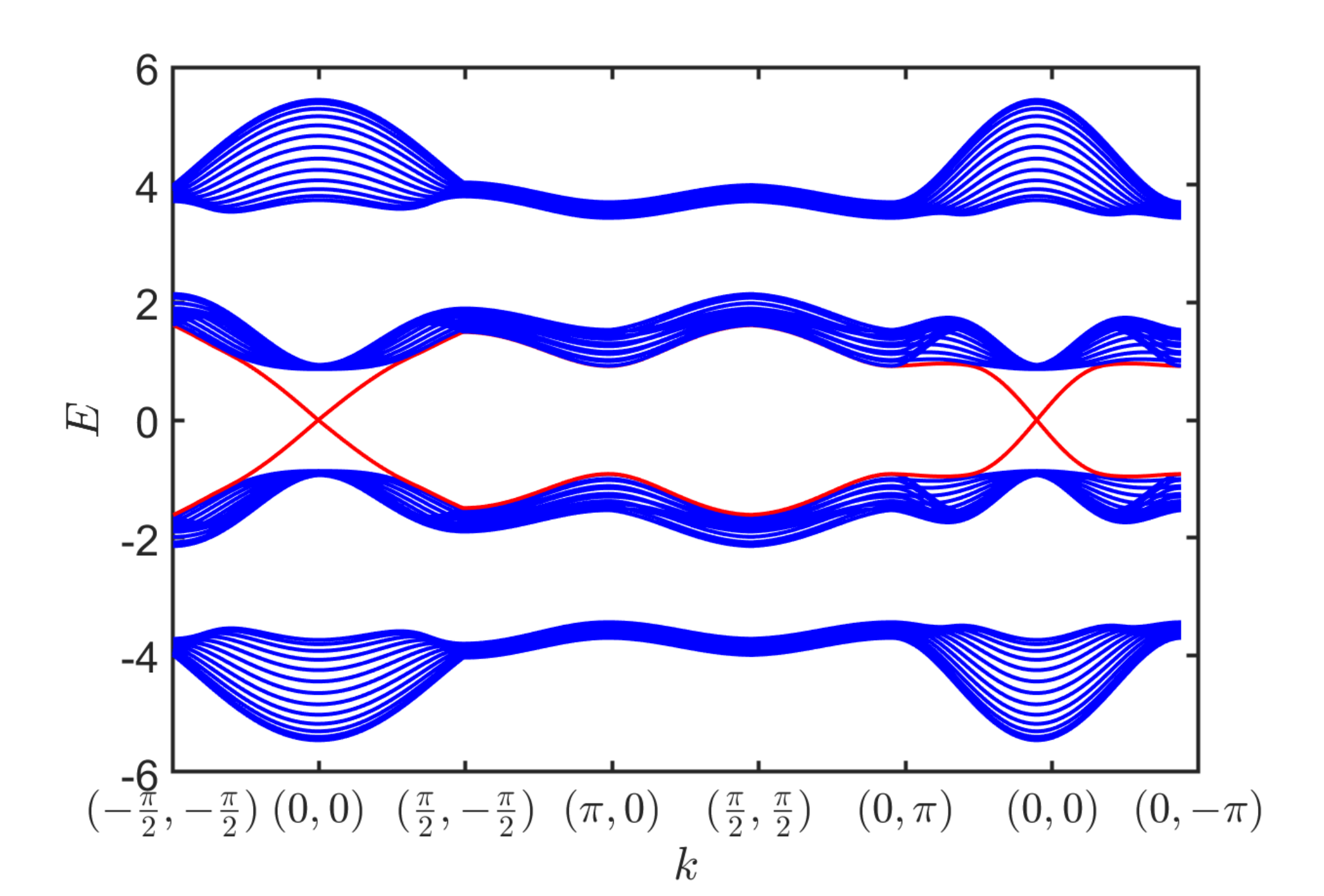}
        \caption{}
    \end{subfigure}
    \begin{subfigure}[b]{0.4\textwidth}
        \centering        
        \includegraphics[width=\textwidth]{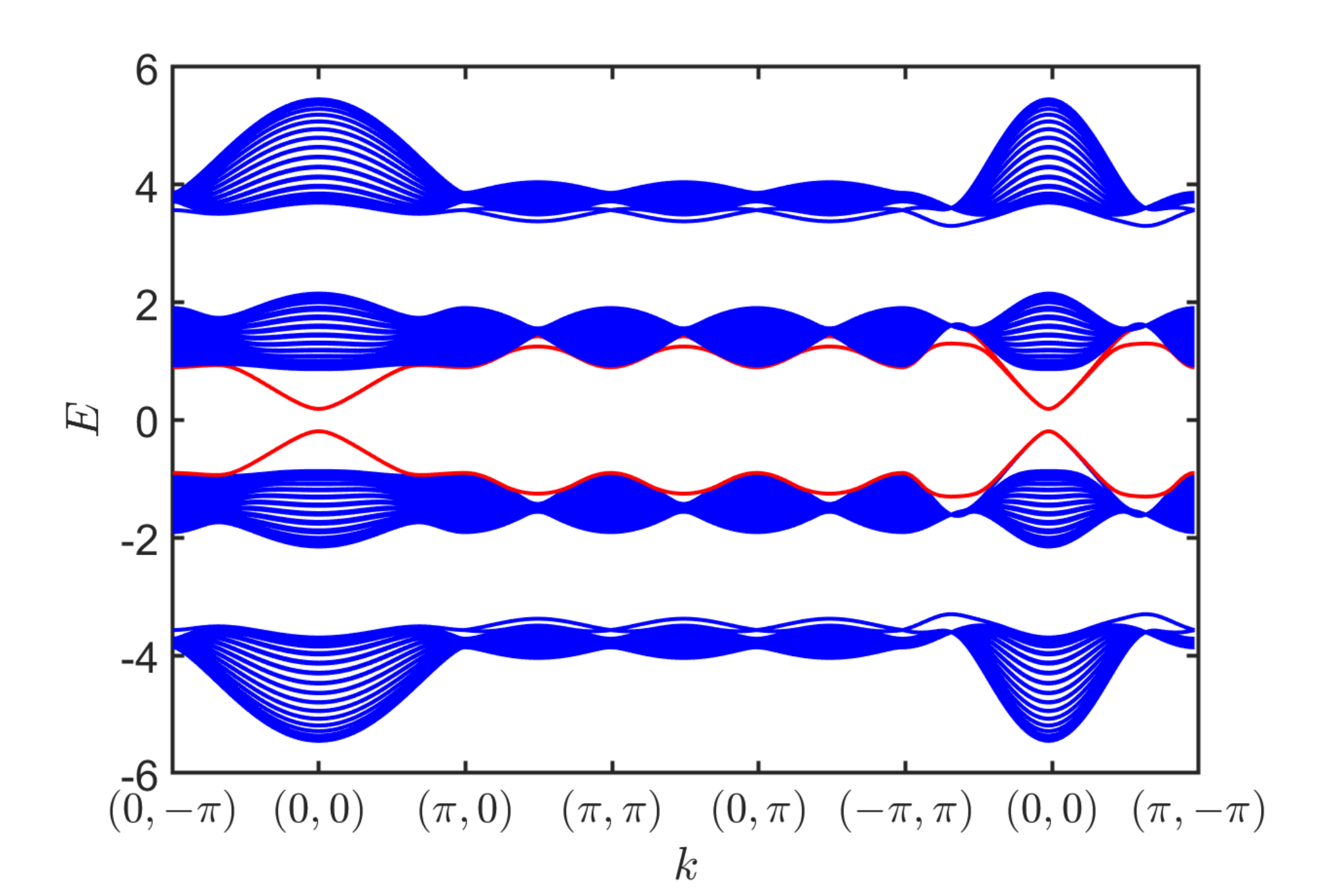}
        \caption{}
    \end{subfigure}
\caption{  
(a)The topological phase diagram for renormalized mean-field theory with Dresselhaus interaction. Here we set $\bar{v}_F=1$.  The antiferromagnetic order is in $\{1,1,1\}$ direction and is set to be $(M_{AF,x}, M_{AF,y}, M_{AF,z}) = (0.577,0.577,0.577)$ and (b)The topological phase diagram for Rashba interaction with antiferromagnetic order being in $\{1,1,0\}$ direction. Here we can find the mean-field solution only in the red rectangular region with $\bar{t} \geq 0.019$, and we choose $(M_{AF,x}, M_{AF,y}, M_{AF,z}) = (0.707,0.707,0)$ and $\bar{v}_F=1$ to illustrate the phase diagram.  Here in (a) and (b), the topological index based on the $S$-symmetry is given by $Z^M_2$; while for the same parameter, it can be also characterized by the $Z_4$ index. Clearly, we see that $Z^M_2$ and $Z_4$ turns out to be the same, and $M_{AF,z}$ is irrelevant so that two phase diagrams are identical.   Note that for a given ratio of $t/m$, topological phases that the system can have when doping increases from $0$ to finite $\delta$ are phases passing by the line with slope $t/m$ on the phase diagram.
(c)The computed $S$-symmetry protected surface state in $\{1,0,0\}$ surface for Dresselhaus interaction with antiferromagnetic order being in $\{1,1,1\}$ direction in $13$ layers. Here $(M_{AF,x}, M_{AF,y}, M_{AF,z}) = (0.577,0.577,0.577)$. Similar structure of the surface state is also found for Rashba interaction with antiferromagnetic order being in $\{1,1,0\}$ direction with $(M_{AF,x}, M_{AF,y}, M_{AF,z}) = (0.707,0.707,0)$.(d) For both models computed in (c), the surface states become gapped when the effective time-reversal symmetry ($S$-symmetry) is broken in the $\{\Bar{1},\Bar{1},1\}$ surface. Here in the $\{\Bar{1},\Bar{1},1\}$ surface, the magnetic order is ferromagnetic and the computation is done in $31$ layers.  }
\label{fig:anti_Z2}
\end{figure}
\end{center}

Given any antiferromagnetic order  $\vec{M}_{AF}$ that may not be mean-field solutions, one can follow the above method to find the corresponding topological indices for renormalized mean-field Hamiltonian.
In particular, to determine topological indices $Z_2$ and $Z_4$,  only parameters $\bar{t}$ and $\bar{m}$ in $H^{D/R}_M$ are relevant. Hence topological phase diagram is presented in the parameter space $\bar{t}$-$\bar{m}$. Furthermore, because $\bar{t}$ and $\bar{m}$ depends on doping (cf. Eq. \ref{bar}) and $\bar{t}/\bar{m}= t/m$, for a given ratio of $t/m$, topological phases that the system can have when doping increases from $0$ to finite $\delta$ are phases passing by the line with slope $t/m$ on the phase diagram.
Now based the above  analysis on $Z_2$ and $Z_4$, the $S$-symmetry (effective time-reversal symmetry) protected phase diagram of two models are shown in Figs.\ref{fig:anti_Z2} (a) and (b) with antiferromagnetic order being in $\{1,1,1\}$ and $\{1,1,0\}$ direction respectively. Here we choose $(M_{AF,x}, M_{AF,x}, M_{AF,x}) = (0.577,0.577,0.577)$ and $\bar{v}_F=1$ in Fig.\ref{fig:anti_Z2}(a); while $(M_{AF,x}, M_{AF,x}, M_{AF,x}) = (0.707,0.707,0)$ and $\bar{v}_F=1$ are chosen in Fig.\ref{fig:anti_Z2} (b).  Clearly, it is seen that in this case, topological indices are independent of the form of spin-orbit interactions so that two phase diagrams are of the same. 
Note that we can find the mean-field solution only in the red rectangular region with $\bar{t} \geq 0.019$ and the phase diagram is independent of directions of $M_{AF}$. The corresponding $Z_4$ index is computed based on Eq.(\ref{Z4_index}) through time-reversal invariant momenta (TRIM) in the Brillouin zone defined by the Bravais lattice vectors, $a_1=(1,0,1)$, $a_2=(0,1,1)$, and $a_3=(0,0,2)$ Here reciprocal lattice vectors are given by $g_1 = \pi(1, -1, 1), g_2 = \pi(1, 1, -1)$, and $g_3 = \pi(-1, 1, 1)$ and TRIMs are $\kappa=\frac{n_1}{2}g_1+\frac{n_2}{2}g_2+\frac{n_3}{2}g_3$ with $n_1,n_2,n_3=0, 1$. Furthermore,  we see that $Z^M_2$ and $Z_4$ turns out to be the same, and $M_{AF,z}$ is irrelevant so that two phase diagrams are identical. 
From  Fig.\ref{fig:anti_Z2} (a) and (b), one sees that when $\bar{t}=1$ and $1.415\leq \bar{m} \leq  3.162$, the system is in topological non-trivial phase. To investigate the corresponding surface states, we note that in the  Mong and Moore's approach, $Z^M_2$ index is associated with the $k_z=0$ plane. Hence surface states that are protected are  in the $\{1,0,0\}$ and $\{0,1,0\}$ surfaces. However, as shown in Fig.\ref{fig:anti_Z2} (a) and (b),  the phase region described by non-trivial $Z^M_2$ index are also characterized by non-trivial $Z_4$. In fact, this topological phase is characterized by $(0,0,0;2)$ in terms of the $(Z_2)^3 Z_4$ index. Hence gapless surface states are also present in any surface that preserve the $S$-symmetry, such as $\{0,0,1\}$ surface. To illustrate properties of the surface states, we set parameters to $\bar{v}_{F}=1, \bar{t}=1, \bar{m}=2.3$.  In Fig.\ref{fig:anti_Z2}(c), we show the calculated surface state in $\{1,0,0\}$ surface for  Dresselhaus interaction with antiferromagnetic order being in ${1, 1, 1}$ direction. Here total number of layers in the computation is $31$ and  $(M_{AF,x},M_{AF,y},M_{AF,z}) = (0.577, 0.577, 0.577)$. Similar surface structure is also found for Dresselhaus interaction with antiferromagnetic field being in ${1, 1, 0}$ direction with $(M_{AF,x},M_{AF,y},M_{AF,z}) = (0.707, 0.707, 0)$.
On the other hand, a gaped surface state arises in $\{1,1,1\}$ surface due to its being a ferromagnetic surface, see Fig.\ref{fig:anti_Z2}(d). 
Note that here in order to sketch the surface state, we have performed the necessary transformation by rewriting the Hamiltonian using the Bravais lattice vectors, $a_1=(1,0,1)$, $a_2=(0,1,1)$, and $a_3=(0,0,2)$, and in particular, the $\{\Bar{1},\Bar{1},1\}$ surface is parallel to $a_1$ and $a_2$. These results would correspond to results obtained in the previous research\cite{Mong_Moore}. 

\onecolumngrid

\begin{center}
\begin{figure}[h]
    \begin{subfigure}[b]{0.32\textwidth}
        \centering        
        \includegraphics[width=\textwidth]{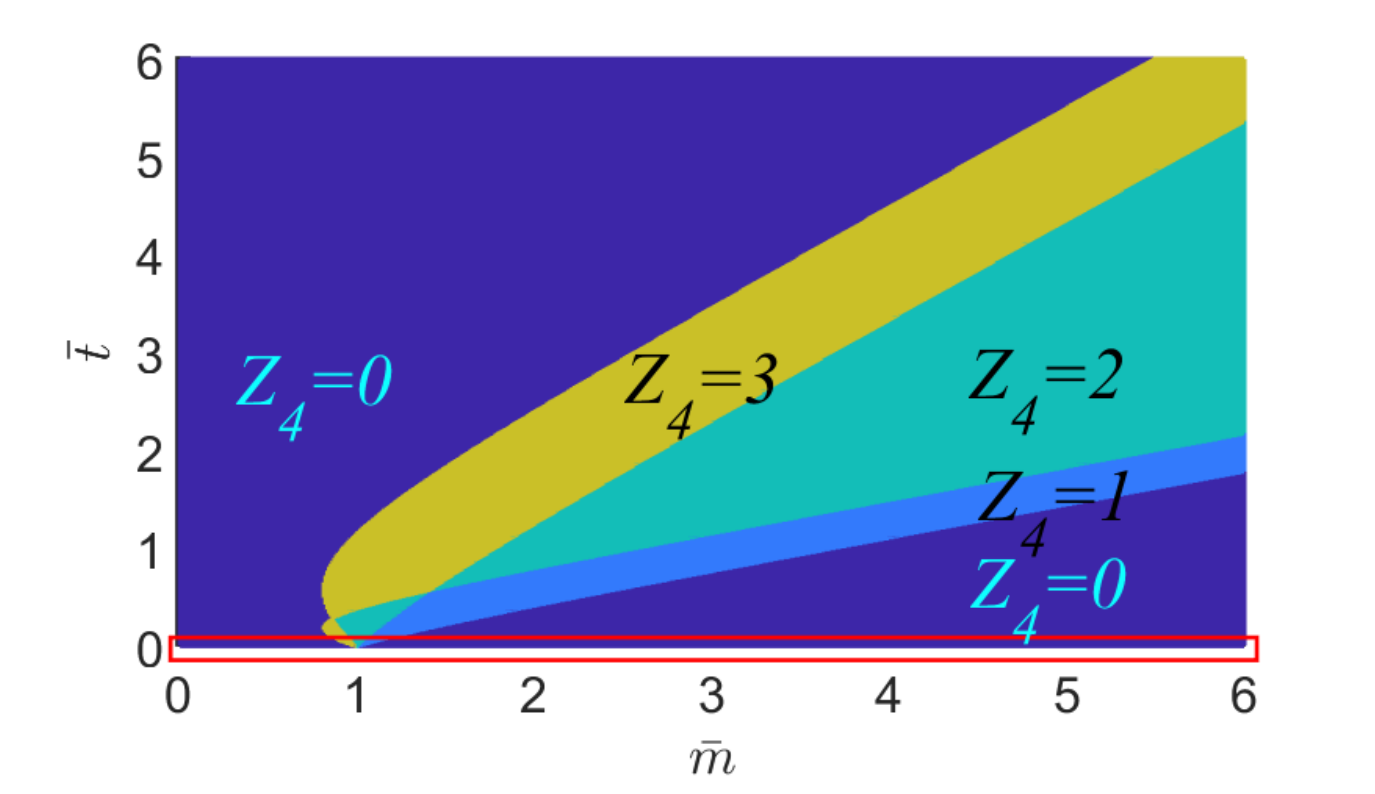}
        \caption{}
    \end{subfigure}
    \begin{subfigure}[b]{0.32\textwidth}
        \centering        
        \includegraphics[width=\textwidth]{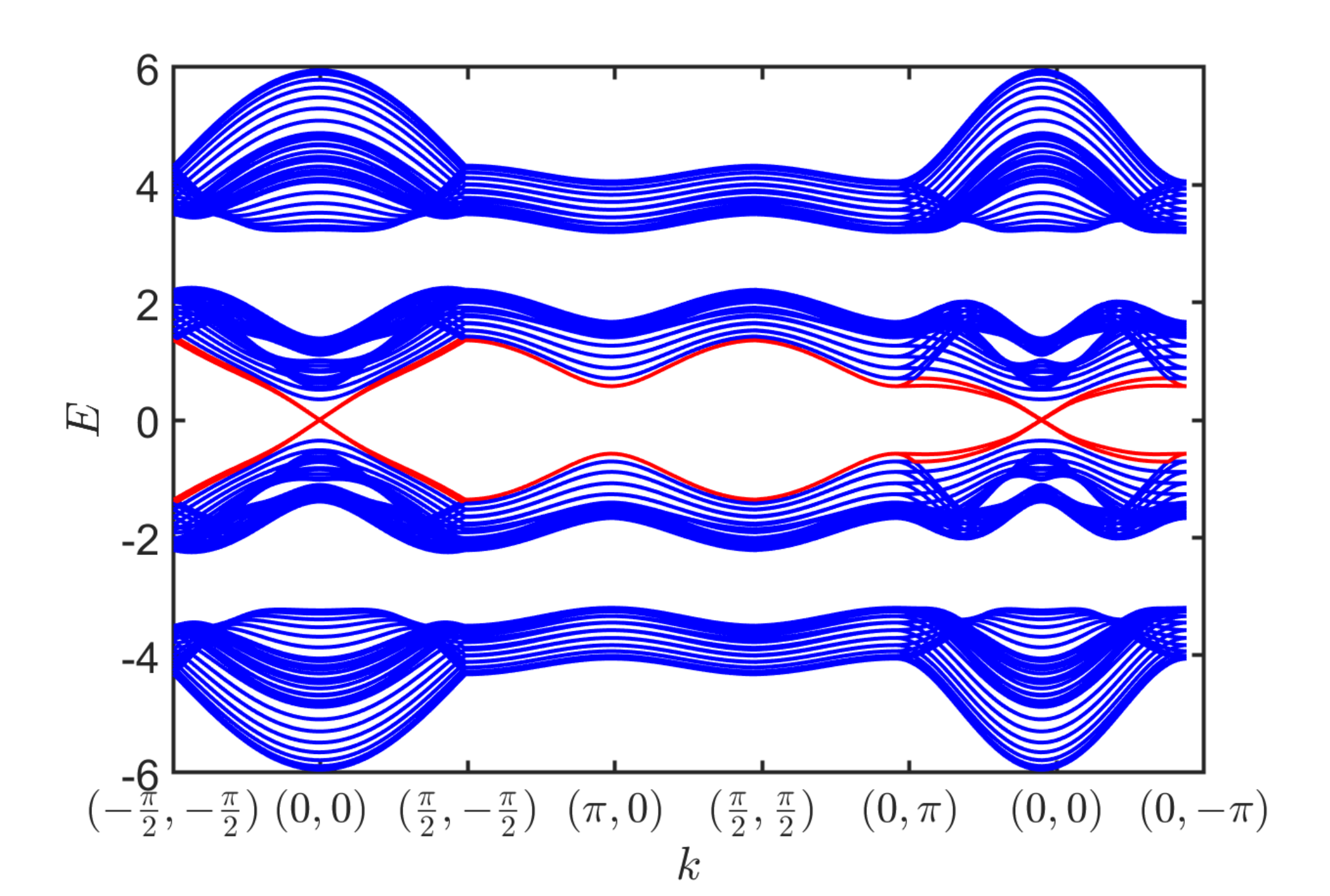}
        \caption{}
    \end{subfigure}
    \begin{subfigure}[b]{0.32\textwidth}
        \centering        
        \includegraphics[width=\textwidth]{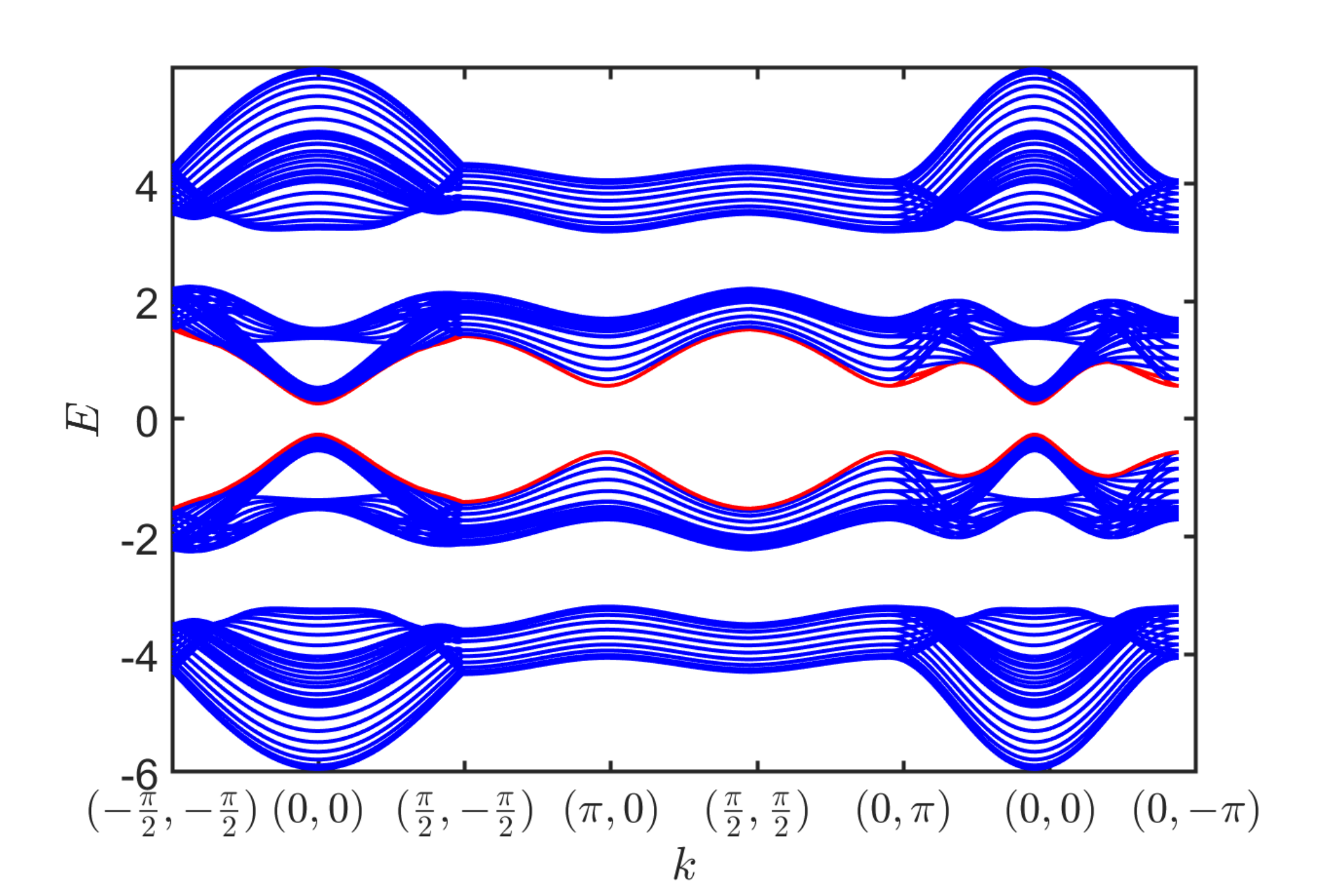}
        \caption{}
    \end{subfigure}    
    
    \begin{subfigure}[b]{0.32\textwidth}
        \centering        
        \includegraphics[width=\textwidth]{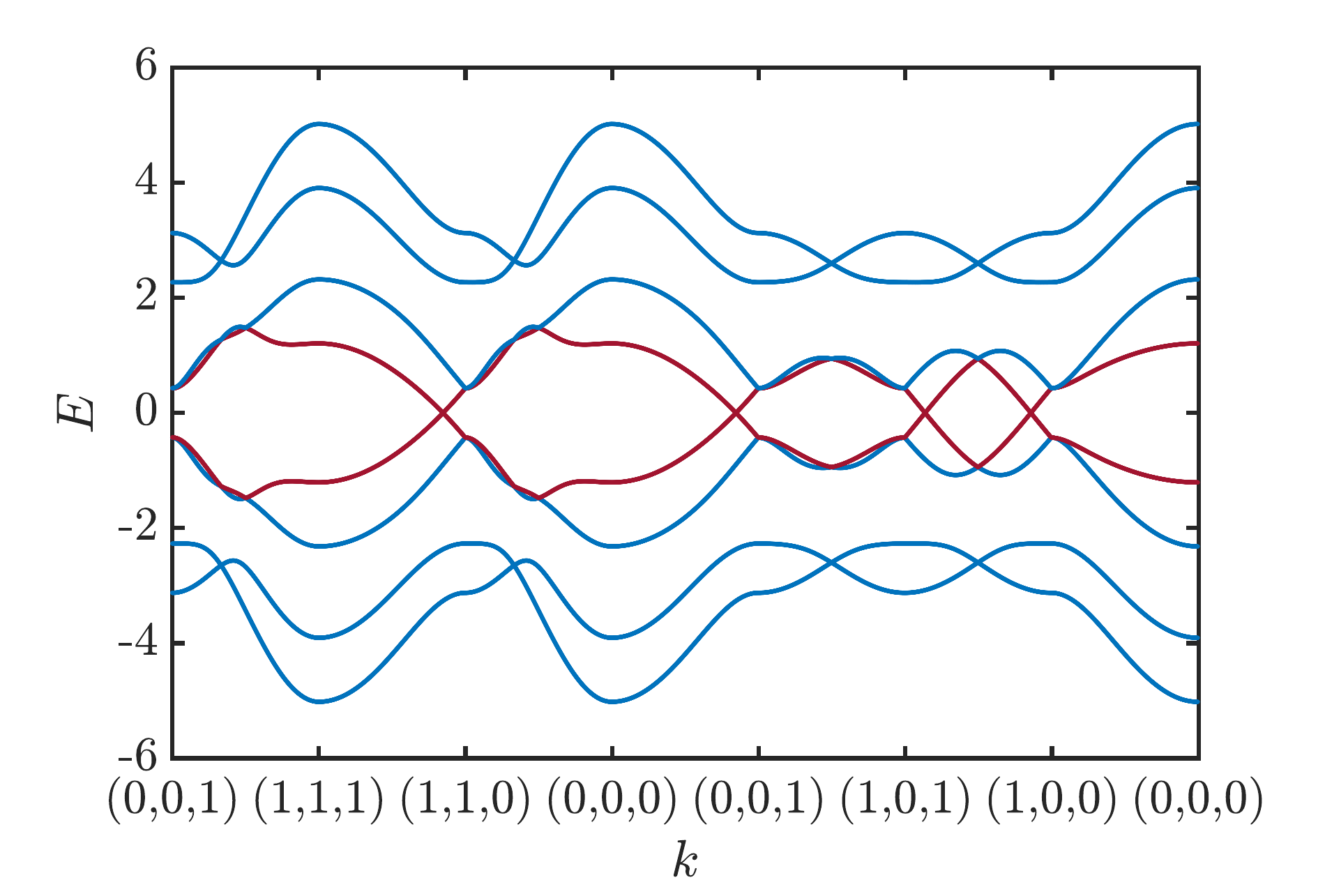}
        \caption{}
    \end{subfigure}
    \begin{subfigure}[b]{0.32\textwidth}
        \centering        
        \includegraphics[width=\textwidth]{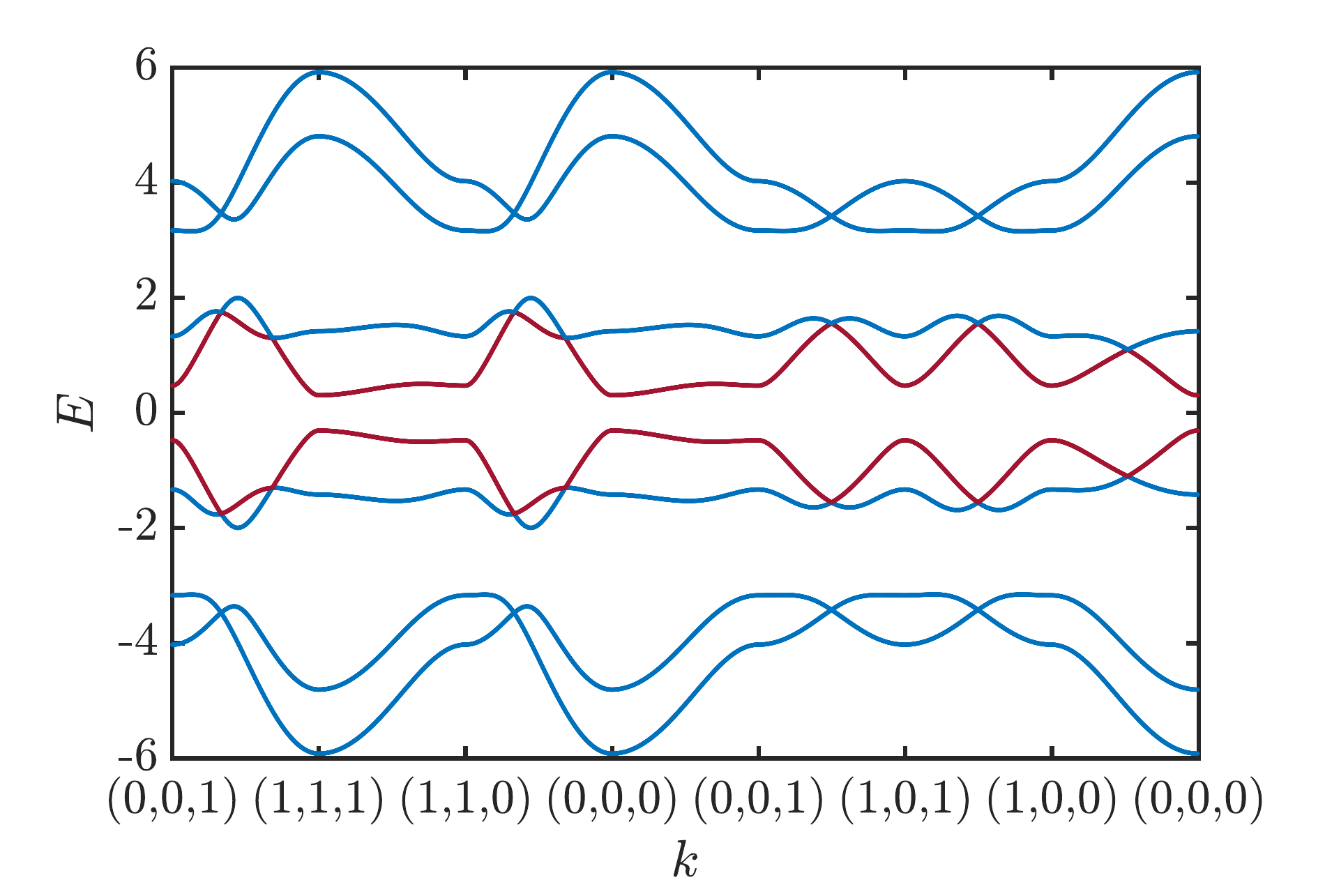}
        \caption{}
    \end{subfigure}
    \begin{subfigure}[b]{0.32\textwidth}
        \centering        
        \includegraphics[width=\textwidth]{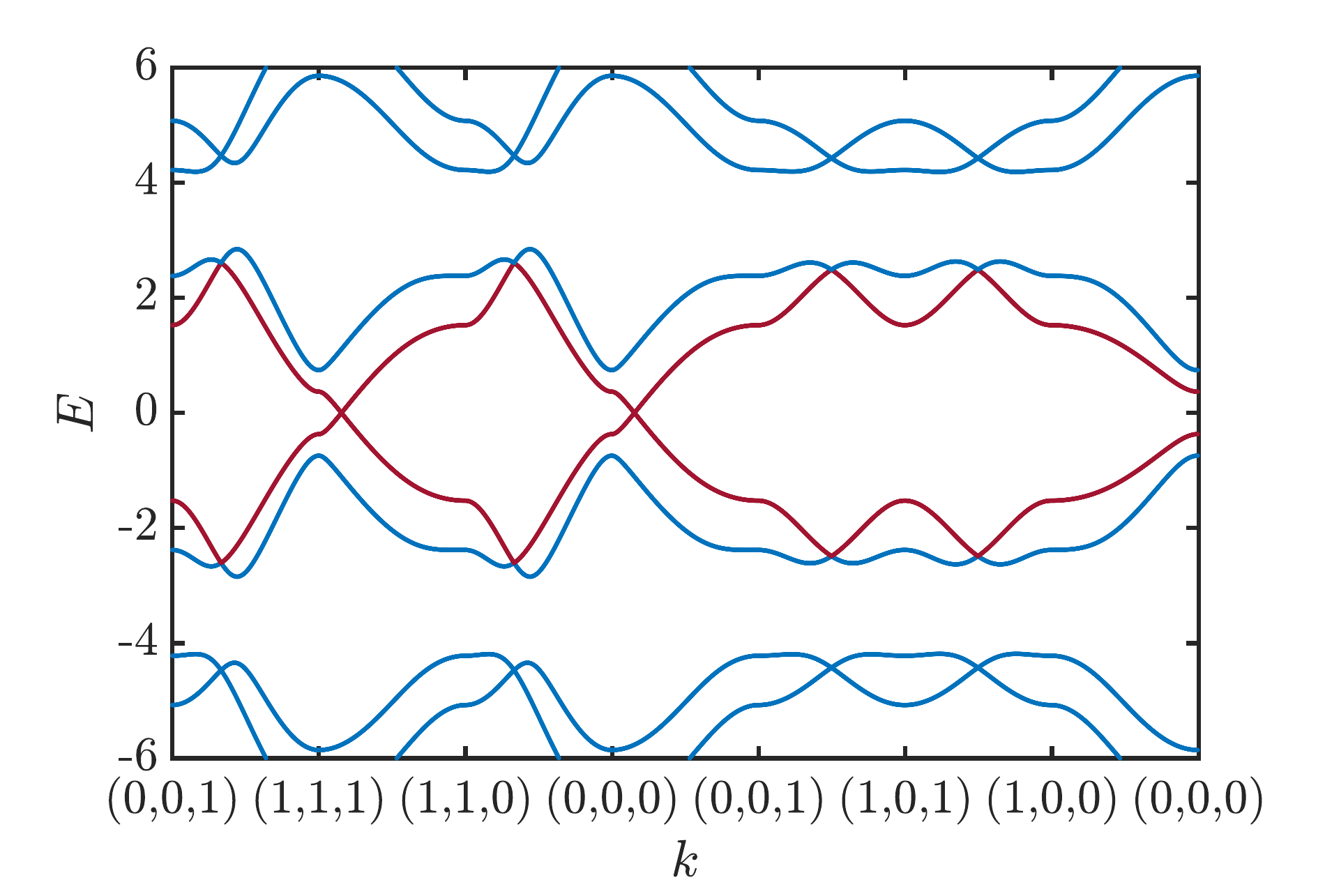}
        \caption{}
    \end{subfigure}
\caption{(a)The $Z_{4}$ phase diagram for Rashba interaction with antiferromagnetic order being in $\{1,1,0\}$ direction and ferromagnetic field in z direction.  Here we can find mean-field solution only in the red rectangular region with  $\bar{t} \leq 0.019$.  Furthermore, for a given ratio of $t/m$, topological phases that the system can have when doping increases from $0$ to finite $\delta$ are phases passing by the line with slope $t/m$ on the phase diagram. (b)The gapless surface state in $\{1,0,0\}$ surface and (c) the gapped surface state in $\{0,0,1\}$ surface with $\bar{m}/\bar{t}=2.3$. Here the computation is done in $13$ layers. The bulk band structure in semi-metallic phase (d)(f) and topological non-trivial phase (e). Here $\bar{t}=1$ and $\bar{v}_F=1$.  (d)$\bar{m}=1.35$, $Z_4=1$. (e) $\bar{m}=2.25$, $Z_4=2$. (f)$\bar{m}=3.3$ and $Z_4=3$.}
\label{fig:anti_ferro_rashba_mxy_mz_Z4}
\end{figure}
\end{center}

\twocolumngrid
We further investigate topological phases for the case when the magnetic order is antiferromagnetic in $x$-$y$ plane but is ferromagnetic in $z$ direction. This is the phase described in Fig. \ref{fig:order diagram}(g).  Here because the Hamiltonian includes the Rashba interaction with non-zero $M_{F,z}$ and the system no longer preserves the effective time-reversal symmetry,  the topology is now protected by the inversion symmetry.  To investigate topological phases associated with a given magnetic order, we take $(M_{AF,x}, M_{AF,x}) = (0.577,0.577)$, $M_{F,x}=0.577$ and $\bar{v}_F=1$. The corresponding topological phases specified by the $Z_4$ index are shown in Fig. \ref{fig:anti_ferro_rashba_mxy_mz_Z4} (a). Note that here we can find the mean-field solution only in the red rectangular region with  $\bar{t} \leq 0.019$. To illustrate surface states and bulk electronic structure, we set $\bar{t}=1$. In this case, the system is in the semi-metallic phase for $0.914\leq \bar{m} \leq 1.764$ and $2.561 \leq \bar{m} \leq 3.659$, while the system is in topological non-trivial phase for $1.765\leq \bar{m} \leq 2.56$, in consistent with Fig. \ref{fig:anti_ferro_rashba_mxy_mz_Z4} (a).  In Figs. \ref{fig:anti_ferro_rashba_mxy_mz_Z4} (d), (e), and (f), we examine the bulk band structure in semi-metallic phase region (d)(f) and topological non-trivial phase region (e). It is clear that the energy gaps exhibited are consistent with the corresponding phases. Finally, in Fig.\ref{fig:anti_ferro_rashba_mxy_mz_Z4}(b) and (c),  we check the surface states for the topological non-trivial phase with $\bar{m}=2.3$. Clearly, there is a gapless point in $\{1,0,0\}$ surface and a gapped surface state in $\{0,0,1\}$ due to the presence of $z$-component of ferromagnetic order and the preservation of the inversion symmetry\cite{Ono_Z4}. In this case, for $\{1,0,0\}$ surface, the inversion symmetry is kept. hence it hosts the gapless surface Dirac modes; while for $\{0,0,1\}$ surface, it breaks the inversion symmetry, resulting in a gapped surface state. How the surface state in $\{1,0,0\}$ becomes gapped can be also understood through alternating stacking of Chern insulators with the $+1$ and $-1$ along $z$ direction. We refer the reader to Ref.\cite{Ono_Z4} for more details.

\onecolumngrid

\twocolumngrid

\section{\label{sec:}DISCUSSION AND CONCLUSION}
In summary, in this paper, we computed topological electronic structures of non-collinear magnetic phases in a multi-orbital Hubbard model with spin-orbit interactions.
By employing self-consistent calculations on magnetic orders, we find distinctive spin arrangements under Dresselhaus or Rashba spin-orbit interactions.
In particular, we find that in the case of the Dresselhaus spin-orbit interaction, the system is collinear antiferromagnetic insulator; while in the case of the Rashba spin-orbit interaction
with hopping amplitude $\bar{t} \leq0.019$, spins are antiferromagnetic in  $xy$-plane but are ferromagnetic aligned in $z$.
We categorize topological properties of these spin arrangements. For given magnetic orders, we  categorize the
topological phases in hopping $t$ and mass $m$ space. Specifically,  for 3D collinear antiferromagnetic order, we find that the system possesses a modified time-reversal symmetry, characterized by the Z$_2$ index. In contrast, for systems with tilted antiferromagnetic orders, we show that the non-trivial topology is protected by the inversion symmetry and is characterized by the Z$_4$ index. Moreover, we also examine the bulk-edge correspondence for non-collinear magnetic phases, we find that indeed, for surfaces that break bulk
symmetry, the corresponding Dirac surface states are gapped. In the case of tilted antiferromagnetic orders we find, this is reflected in the results that
the surface state becomes gapless when the surface is perpendicular to the ferromagnetic component of tilted antiferromagnetic order; otherwise, the surface state exhibits a gap. 
While our results are based on mean-field theory of magnetic orders, we expect that when fluctuations of spin are included, magnetic orders and exchange coupling constants will be normalized and the phase boundaries are renormalized and are slight changed. Hence for a given magnetic order, the overall global picture of topological phases would be 
qualitatively correct. Thus our findings offer a comprehensive topological characterization for doped and canted antiferromagnetic insulators with spin-orbit interactions and provide valuable insights into the interplay between spin arrangements, symmetries, and topological properties in these systems. In particular, recently, there has been intensive researches on the realization of Kitaev model\cite{Kitaevmodel_kim2021antiferromagnetic, Kitaevmodel_sears2020ferromagnetic, Kitaevmodel_shangguan2023one, Kitaevmodel_zhang2022theory}. The Mott insulator with spin-orbit interaction is one of the candidates.  With appropriate generalization, our analysis on multi-orbital Hubbard model with spin-orbit interactions would be a useful access for exploration of topological phases in doped Kitaev materials.

\begin{acknowledgments}
This work was supported by the National Science and
Technology Council (NSTC), Taiwan. We also acknowledge
support from the Center for Quantum Science and
Technology (CQST) within the framework of the Higher
Education Sprout Project by the Ministry of Education
(MOE) in Taiwan.
\end{acknowledgments}


\end{document}